\documentclass[pre,aps,showpacs,12pt,reprint,superscriptaddress,floatfix]{revtex4-1}
\usepackage[mathletters]{ucs}
\usepackage[utf8]{inputenc}
\usepackage{mathtools} 
\usepackage{graphicx}
\usepackage{amsmath}
\usepackage{amsfonts}
\usepackage{amssymb,yfonts,bbm}
\usepackage[bbgreekl]{mathbbol}
\usepackage{mathrsfs}   
\usepackage{color}					
\usepackage{bm}      				
\usepackage{chemarrow}				
\usepackage{booktabs}				
\usepackage{array}					
\usepackage[bbgreekl]{mathbbol}  	
\usepackage{slashed}  				
\usepackage[hidelinks]{hyperref}	
\hypersetup{colorlinks=true,citecolor=blue,urlcolor=blue} 
\usepackage{tikz}					
\usepackage{verbatim}
\usepackage{notes2bib}				
\bibnotesetup{note-name =,use-sort-key = false}

\DeclareMathAlphabet{\mathpzc}{OT1}{pzc}{m}{it}
\usepackage{svg}        
\usepackage{enumitem}
\usepackage{xcolor}

\newtagform{hp}{\, (HP\;}{)}
\newtagform{eg}{\, (EG\;}{)}
\newtagform{hhp}{\, (HHP\;}{)}
\newtagform{chp}{\, (CHP\;}{)}
\newtagform{eg,hhp}{\, (EG,\;HHP\;}{)}

\begin{document}
\title{Thermodynamic Circuits: Association of thermoelectric converters in stationary non-equilibrium}

\author{Paul Raux}
\affiliation{Université Paris-Saclay, CNRS/IN2P3, IJCLab, 91405 Orsay, France }
\affiliation{Université Paris Cité, CNRS, LIED, F-75013 Paris, France}
\author{Christophe Goupil}
\affiliation{Université Paris Cité, CNRS, LIED, F-75013 Paris, France}
\author{Gatien Verley}
\affiliation{Université Paris-Saclay, CNRS/IN2P3, IJCLab, 91405 Orsay, France }

\date{\today}

\begin{abstract}
Following up on the recently published circuit theory for thermodynamic devices, we consider networks of Thermo-Electric Converters (TECs) in stationary non-equilibrium. Assuming constant thermoelectric properties, the integration over a finite thickness of the linear local response of the thermoelectric material yields the non-linear current-force characteristics. We show how to derive a choice of nonequilibrium conductance matrix summarizing the current-force characteristics for every available sets of currents and forces. This problem has infinitely many solutions if one considers only thermodynamic constraints. Each solution differs, among others, by the coupling between the currents. Then, we determine the current-force characteristics of the serial (respectively parallel) association of two TECs using the laws of resistance (respectively conductance) matrix addition. For TECs in series, we find current-dependent boundary conditions for each sub-device. Since currents derive from composite potentials, we also associate the derivability and continuity of these potentials at the interfaces with conditions on thermoelectric coefficients. For TECs in parallel, we discuss the possibility of loop currents that are forbidden for the serial association.
\end{abstract}

\maketitle

\newpage

\section*{Introduction}

Investigations on the connection of thermodynamic devices in networks started in the first half of the last century~\cite{Michaelis1930, Teorell1953} and experienced later a strong development, mainly in biology and bioenergetics~\cite{katchalsky1965nonequilibrium, Oster_Perelson_Katchalsky_1973}. 
In particular, the work of O. Kedem and R. S. Caplan made it possible to describe coupled transport phenomena across biological membranes under nonequilibrium conditions~\cite{Kedem1965}. Historically, these questions of coupled transport in biology were quickly associated with those of the chemical reactions that govern biochemistry~\cite{KRAMERS1940}. Beyond the field of biology, a current-force approach emerges at the core of irreversible thermodynamics~\cite{Schnakenberg1976, Hill1989, Peusner1985, Broeck2015}. Thermoelectric conversion is a paradigmatic application of this framework in the weakly irreversible limit~\cite{Onsager1931}. 
Indeed, in 1948, Callen proposed a concise model for describing TEC requiring only three parameters called thermoelectric coefficients~\cite{Callen1948}. This phenomenological model, called Constant Property Model, allowed for an analytical study of the energy and matter transport and their coupling. At the mesoscopic scale, these parameters remain constant irrespective of thermodynamic forces. This modeling approach was extended in 1959 for thermoelectric material of macroscopic thickness by A. Ioffe to incorporate explicit dissipative contribution and entropy balances with thermostats~\cite{Ioffe1959}. Nowadays, the characteristic equations derived by Ioffe still prevail to design efficient TEC, either dithermic engines producing electric power or, in reverse operating mode, to pump heat against thermal gradient at the expense of electric work~\cite{Apertet2013, Feldhoff2020, Raux2025}.

This manuscript includes two main parts with different purposes~\footnote{During the review process, referees demanded to merge the second and third articles of this series compared to what is announced in Ref.~\cite{Raux2024}.}: 
First, we focus on defining the notion of nonequilibrium conductance matrix for a TEC without assuming any underlying microscopic dynamics~\cite{Esposito2015, Rax2015}. Second, we build on these results to study the association of TECs made with different (or identical) materials. 

In the first part, we demonstrate that the non-linear current-force characteristics obtained by A. Ioffe can be cast in matrix form. To do so, we derive the characteristic equations based on composite potentials (thermal and electric) and exhibit a choice of compatible conductance matrix: The corresponding freedom arises from the choice of currents coupling \cite{Vroylandt2018}, similar to the figure of merit in the linear regime \cite{Kedem1965}. Alternatively, a microscopic dynamics would prescribe the current coupling far from equilibrium, as it does in the linear regime. Such a microscopic dynamics can predict as well the current fluctuations that are only bounded by the thermodynamically consistent modeling based on nonequilibrium conductance matrices~\cite{Vroylandt2018,Vroylandt2019}. In any case, our modeling allows arbitrary degrees of coupling, improving upon the notion of strong coupling (i.e., proportional currents with the same proportionality factor for all forces) while remaining at the level of irreversible thermodynamics. In addition, in the context of energy conversion, it is relevant to express the conductance matrix for the currents inter-converted. Those are different from the conserved currents at the core of our circuit theory. Therefore, we express the conductance matrix in different bases of currents, emphasizing the subtle convective and conductive nature of energy transport.

The second part of this article focuses on the association of two TECs based on the modeling introduced in the first part. We determine the effective descriptions for the serial and the parallel association of two TECs by applying respectively the law of resistance matrix addition and the law of conductance matrix addition~\cite{Raux2024}. Consequently, for a given coupling of charge and energy currents, i.e., for a given choice of conductance/resistance matrices, we expect to obtain a unique conductance matrix for the composite device, and hence the coupling between the currents.
We show in particular that under fixed temperature difference, the voltage-current characteristics of the serial association of two TECs (with linear voltage-current characteristics) is nonlinear. 
This non linearity is due to the current dependent boundary conditions (intermediate conditions between Dirichlet and Neumann). This situation is typical and also occurs, for example, when modeling the force-velocity response of muscles in animal locomotion~\cite{Goupil2019}. Finally, we show that the material property mismatch leads to interface heat exchange due to the Peltier effect: temperature and electric potentials are continuous by assumption, whereas the composite potentials introduced in the first part are discontinuous in case of mismatch.
This interface problem led in particular to the concept of relative current~\cite{PhysRevLett.91.148301,Goupil2011} which highlights the challenge of optimally coupling thermodynamic converters of materials with different transport parameters \cite{PhysRevLett.91.148301, Goupil2009}. This point is central in that it directly questions impedance adaptation in the case of nodal structures associating coupled forces and currents of different natures. 

The first part of this article is organized as follows: 
Section~\ref{sec : from local to global} provides an original derivation of the current-force characteristics of a TEC (Ioffe's equations) starting from the local flux-force relation. It showcases how a global-scale framework in thermoelectricity unveils space-invariant quantities, such as the so-called $F$ functions, which characterize the free fraction of energy in the CPM. 
Section~\ref{sec : non equilibrium conductance matrix} focuses on the possible forms of the nonequilibrium conductance matrix. First, we describe its general internal structure: We show that the conductance matrix at the level of fundamental currents (linearly independent) can be read as a sub-block of the conductance matrix at the level of physical currents (linearly dependent currents due to conservation laws). Second, we define and relate external (i.e., exchanged with reservoirs) and internal (i.e., crossing the device) physical currents. Third, we relate those currents to the fundamental ones. This section illustrates the selection of fundamental currents and forces on the concrete case of TEC~\cite{Raux2024}. Interestingly, it also extends our understanding of conservation laws, enabling the linear combination of physical currents with non-integer ($0$ or $1$) values. Restarting directly from Ioffe's equations, section \ref{CondMatrix} provides a conductance matrix for fundamental currents and the corresponding one for physical currents, either internal or external. 

The second part of this article is organized as follows: 
In section \ref{sec : serial association}, we derive the nonequilibrium conductance matrix describing the serial association of two TECs by applying the law of resistance matrix addition~\cite{Raux2024}. 
In section \ref{sec : parallel association}, we considers the parallel association of two TEC in Dirichlet boundary conditions (pins of both TECs at the same fixed potentials). We show that in general, the mismatch between the thermoelectric coefficients yields in this case internal currents even in open circuits configurations (no external currents from the reservoirs).

\section{Nonequilibrium conductance matrix for a TEC}
\label{DefG}

\subsection{Current-force characteristics of the CPM}
\label{sec : from local to global}

In this section, we start from the local (and linear) description of a TEC in terms of a flux-force characteristics defined by the Onsager matrix and its associated thermoelectric coefficients. Using the conservation of matter and energy fluxes across a surface $\mathcal{A}$ of an homogeneous thermoelectric material of finite thickness $\Delta x$, we obtain the global current-force characteristic. As compared to Ref.~\cite{Apertet2013}, our derivation emphasizes that both charge and energy currents derive from potential functions that we identify. We also shed light on the role of a space independent quantity that we denote $F$. Finally, we rephrase Domenicalli's equation as a necessary condition for the conservation of energy.
\subsubsection{Local Onsager flux-force relation}
Let's start by physically motivating each term in the linear characteristic equations of (an effectively one dimensional) thermoelectric material of infinitesimal thickness $\Delta x$. In such material, the heat flux $J_Q$ (W.m$^{-2}$) and the electrical flux $J_C$ (C.s$^{-1}$.m$^{-2}$) are coupled by the local flux-force relation:
\begin{equation}
\begin{pmatrix}
J_Q\\
J_C 
\end{pmatrix}= \bm L \begin{pmatrix}
-\frac{dT}{dx}\\
-\frac{dV}{dx} 
\end{pmatrix},
\label{eq : electrical heat flux force relation thermoelectric coefficients}
\end{equation}
where the thermodynamic forces are respectively the temperature and the electric potential gradients. The temperature $T=T(x)$ and the electric potential $V=V(x)$ are assumed to be constant in any transverse plane of the material with constant $x$ value. The response matrix appearing in Eq.~\eqref{eq : electrical heat flux force relation thermoelectric coefficients} is
\begin{equation}
\bm L=\begin{bmatrix}
\alpha^2 \sigma_T T + \kappa_J & \alpha \sigma_T T \\
\alpha \sigma_T & \sigma_T \\
\end{bmatrix}, \label{HeatResponseMatrix}
\end{equation}
where $\kappa_J$ is the thermal conductivity under zero electrical current and $\sigma_T$ the isothermal electrical conductivity. Indeed, we recover Fourier's law $J_Q=-\kappa_J\frac{dT}{dx}$ associated to conductive heat flux (respectively Ohm's law $J_C=-\sigma_T \frac{dV}{dx}$) by taking $\alpha \rightarrow 0$ in the first (respectively second) row of Eq.~\eqref{eq : electrical heat flux force relation thermoelectric coefficients}. The Seebeck coefficient $\alpha$ couples heat and charge fluxes and is defined by 
\begin{equation}
	\alpha=-\left. \frac{\frac{dV}{dx}}{\frac{dT}{dx}}\right|_{J_C=0}.
\end{equation}
Dimension analysis accounts for the off-diagonal terms in $\bm L$ that must involve $\alpha$ by definition. The first term $\alpha^{2}\sigma_{T} T$ in the first diagonal component of $\bm L$ arises from convective heat flux. Finally, we remark that the matrix $\bm L$ is not symmetric: Onsager's reciprocity relation does not hold because the force vector $(-\frac{dT}{dx},-\frac{dV}{dx})$ and flux vector $(J_Q,J_C)$ are not conjugated thermodynamic variables. Onsager's reciprocity relations hold in a basis of conjugated thermodynamic variables: here it is the force vector $\left(\frac{d}{dx}\frac{1}{T},-\frac{1}{T}\frac{dV}{dx}  \right)$ that is conjugated to the current vector $(J_Q,J_C)$.

In the following, we integrate the local flux-force relation over a finite thickness of thermoelectric material. To prepare for this integration, it is convenient to introduce the conserved fluxes $(J_E,J_C)$ of energy and charge conjugated to the thermodynamic forces $\left( \frac{d}{dx}\frac{1}{T},-\frac{d}{dx}\frac{V}{T}\right)$. The energy flux $J_{E}$ is defined according to the first law of thermodynamics by
\begin{equation}
J_E=J_Q +V(x)J_C
\label{eq : local first law}
\end{equation}
where $V(x) J_C$ is the flux of electric power crossing the surface $ \mathcal{A}$ (in plane of constant $x$) and oriented toward growing $x$. From the conservation of energy and charge, we remark that the heat flux $J_{Q} = J_E - V(x)J_C$ is $x$ dependent. This is precisely this difference between the heat fluxes across the surface $ \mathcal{A}$ at $x=0$ and at $x=\Delta x$ that allows thermoelectric power generation. In the new basis of flux $(J_E,J_C)$ and forces $\left( \frac{d}{dx}\frac{1}{T},-\frac{d}{dx}\frac{V}{T}\right)$, the local flux-force relation becomes
\begin{equation}
\begin{pmatrix}
        J_E \\
        J_C
    \end{pmatrix}=\bm{ \mathcal{L}} \begin{pmatrix}
    \frac{d}{dx}\frac{1}{T} \\
    -\frac{d}{dx}\frac{V}{T}
\end{pmatrix},
\label{eq : energy matter flux force relation thermoelectric coefficients}
\end{equation} 
involving the response matrix
\begin{equation}
     \bm{ \mathcal{L}}=
    \begin{bmatrix}
        \kappa_J T^2 + T\sigma_T\left( \alpha T + V \right)^2 & T\sigma_T \left( \alpha T +V \right)\\
        T\sigma_T\left(\alpha T +V \right) & T\sigma_T
    \end{bmatrix}. \label{EnergyResponseMatrix}
\end{equation}
This matrix follows from first using in Eqs.~(\ref{eq : electrical heat flux force relation thermoelectric coefficients}--\ref{HeatResponseMatrix}) the following change of variables 
\begin{equation}
\begin{pmatrix}
- \frac{dT}{dx} \\
- \frac{dV}{dx}
\end{pmatrix}=
\begin{bmatrix}
0 & T^{2} \\
VT & T 
\end{bmatrix}
\begin{pmatrix}
\frac{d}{dx}\left(\frac{1}{T}\right) \\
-\frac{d}{dx}\left(\frac{V}{T}\right)
\end{pmatrix},
\end{equation}
and second by using Eq.~\eqref{eq : local first law} to switch from the vector of heat and charge fluxes to the vector of energy and charge fluxes.
In the following, we call Eqs.~(\ref{eq : energy matter flux force relation thermoelectric coefficients}--\ref{EnergyResponseMatrix}) the local flux-force relation for the conserved fluxes. Since we use variables that are conjugated when considering an entropy balance, the matrix $\mathcal{L}$ is symmetric and verifies Onsager's reciprocity relations. It is also $x$ dependent via the local temperature $T$ and electric potential $V$.
\subsubsection{Current-force relation from integrated fluxes \label{integration}}
\begin{figure}[t]
\centering
\includegraphics[width=\columnwidth]{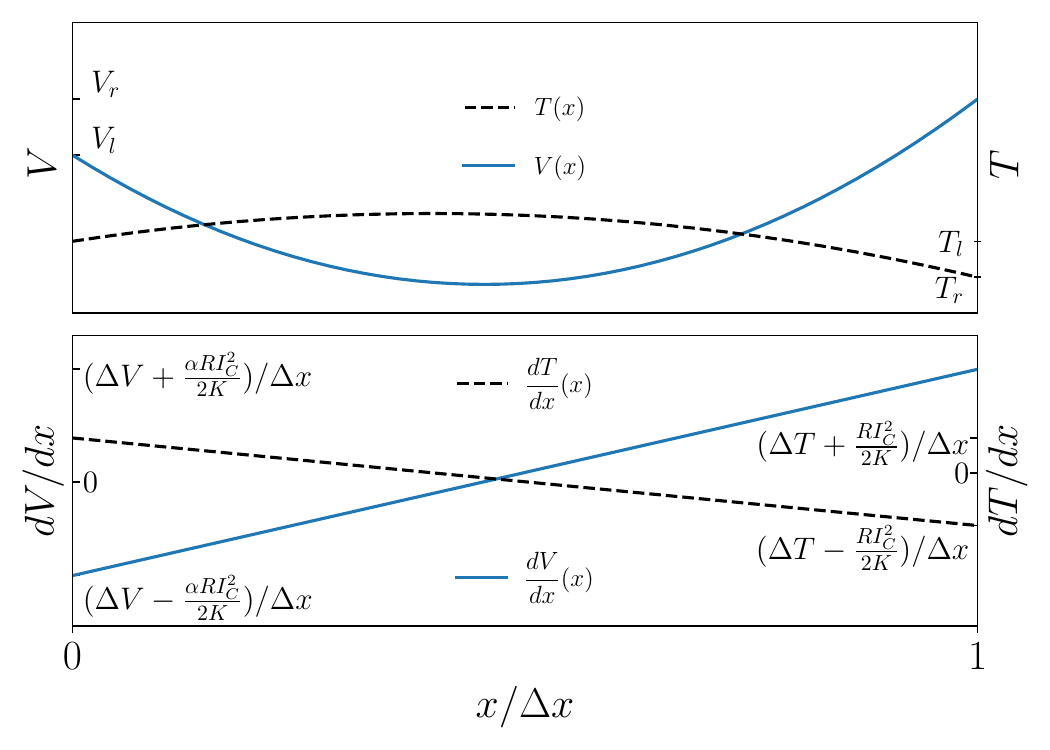}
\caption{Electric potential and temperature profiles (Top panel) and gradients (Bottom panel) along the $x$ axis of the thermoelectric material under the fixed boundary conditions of Eqs.~(\ref{leftBC}--\ref{rightBC}). The left vertical axis is for electric potential $V$ profile (respectively gradients) in blue solid line. The right vertical axis is for temperature profile (respectively gradients) in black dashed line.
\label{fig : schema gradients}}
\end{figure}

In this section, we switch from the local level, with fluxes crossing an infinitesimal slice of thermoelectric material, to the global level, with currents defined as space integrated fluxes over $\mathcal{A}$ and crossing a finite thickness $\Delta x$ of thermoelectric material. This integration step produces a non-linear current-force relation from a linear force-flux relation. When integrating the local flux-force relation, we use Dirichlet boundary conditions for the TEC:
\begin{align}
	T(0)	&=T_l, 	& V(0) 	& =V_l, \label{leftBC}\\
	T(\Delta x)	&= T_r, 	& V(\Delta x) 	&= V_r. \label{rightBC}
\end{align}
Without loss of generality, we assume $T_l>T_r$. By convention, all flux and currents are algebraic and positive when flowing toward the growing $x$ direction. We start by showing that both matter and energy fluxes derive from a potential. Using this property, the global current-force relation is then obtained by integrating the differential equations for those potentials on the finite thickness of thermoelectric material. 
\paragraph*{Electric current}
We start by determining the electric current as a function of the temperature and voltage differences, respectively $\Delta T=T_{r}-T_{l}$ and $\Delta  V=V_r-V_l$, between the right and left planes of the TEC. According to Eqs.~(\ref{eq : electrical heat flux force relation thermoelectric coefficients}--\ref{HeatResponseMatrix}) and for homogeneous materials (constant $\alpha$), the flux $J_C$ of  electric charges derives from a potential $\varphi(x)$ as
\begin{equation}
    J_C = - \sigma_T \frac{d\varphi}{dx},  \label{eq : electric flux}
\end{equation}
where
\begin{equation}
\varphi=\alpha T + V .
\label{eq : def of phi}
\end{equation}
In the stationary state, this flux is divergence-less to ensure charge conservation
\begin{equation}
     \frac{dJ_C}{dx}= - \sigma_T \frac{d^2\varphi}{dx^2}=  0.
    \label{eq : matter conservation}
\end{equation}
From this last equation, 
\begin{equation}
     \frac{d \varphi}{dx} = \frac{d}{dx} \left( \alpha T + V \right)=\mathrm{const}, \label{eq : linear relation dT and dV}
\end{equation}
and $\varphi$ is an affine function of $x$ with coefficients yet to be determined.
For our effectively one dimensional TEC under the boundary conditions of Eqs.~(\ref{leftBC}--\ref{rightBC}), see Fig.~\ref{fig : schema gradients}, the solution of  Eq.~\eqref{eq : matter conservation} reads
\begin{equation}
\varphi(x) = \left( \varphi_r-\varphi_l \right)\frac{x}{\Delta x} + \varphi_l \label{eq : phi (1)}
\end{equation}
where we have denoted $\varphi_\chi=\alpha T_\chi + V_\chi$ with $\chi=l,r$. 

Then, the electric current across the surface $ \mathcal{A} $ follows from the expression of the electric flux \eqref{eq : electric flux} combined with Eq.\eqref{eq : phi (1)} as 
\begin{equation}
I_C\equiv J_C \mathcal{A} =\frac{\varphi_l-\varphi_r}{R}\label{eq : electric current (1)}
\end{equation}
where we have denoted $R=\Delta x/(\sigma_T\mathcal{A})$ the isothermal electric resistance. 
Remarkably, the expression of $I_C$ given in Eq.~\eqref{eq : electric current (1)} takes the form of Ohm's law generalized to the composite potential $\varphi$ which encapsulates the coupling between energy and charge transport. We can finally express $\varphi$ in term of the thermoelectric coefficients and the potentials of the reservoirs as
\begin{equation}
\varphi(x)=  \left(\alpha \Delta T + \Delta V\right) \frac{x}{\Delta x} + \alpha T_l + {V_l}.
\label{eq : solution of matter conservation}
\end{equation}
Then, the first current-force characteristics providing the electric current as function of temperature and voltage differences at the boundaries of the TEC is linear and reads
\begin{equation}
I_C
=-\frac{1}{R}\left(\alpha \Delta T + \Delta V\right).
\label{eq : global matter current}
\end{equation}
We emphasize that in generator convention a positive power is generated by the TEC and absorbed by the load when $I_C\Delta V>0$ (for $\Delta V$ and $I_C$ aligned in generator convention and anti-aligned in load convention).

\paragraph*{Energy current}
We continue with the determination of the energy current as a (non-linear) function of $\Delta T$ and $\Delta V$. According to Eqs.~(\ref{eq : electrical heat flux force relation thermoelectric coefficients}--\ref{HeatResponseMatrix}), the heat flux writes 
\begin{equation}
	J_{Q} = \alpha T J_C - \kappa_{J}\frac{dT}{dx} \label{heatflux}
\end{equation} 
that, when combined with Eq.~\eqref{eq : local first law} and the definition of $\varphi$ in Eq.~\eqref{eq : def of phi}, leads to the energy flux 
\begin{equation}
J_E=\varphi J_C - \kappa_J\frac{dT}{dx}.
\label{eq : je en fonction de j et nabla T}
\end{equation}
Let's show that $J_E$ derives from a potential. Indeed, combining Eq.~\eqref{eq : electric flux} and Eq.~\eqref{eq : je en fonction de j et nabla T} yields
\begin{equation}
J_E= - \kappa_J\frac{d\Phi}{dx},
\label{eq : energy flux}
\end{equation}
where 
\begin{equation}
\Phi= T + \frac{\sigma_T}{2\kappa_J}\varphi^2.
\label{eq : def Phi}
\end{equation}
The potential function $\Phi$ depends on $x$ through $T$ and $\varphi$. Energy conservation is thus ensured if and only if 
\begin{equation}
\frac{dJ_E}{dx}=0=-\kappa_J \frac{d^2\Phi}{dx^2}.
\end{equation}
This last equation is readily solved as
\begin{equation}
\Phi(x)=\left( \Phi_r-\Phi_l \right)\frac{x}{\Delta x} + \Phi_l
\label{eq : Phi(x)}
\end{equation}
where we have denoted $\Phi_\chi=T_\chi + \frac{\varphi_\chi^2}{2KR}$ with $\chi=l,r$. We denote $K=\kappa_J\mathcal{A}/\Delta x $ the thermal conductivity under zero electric current for the TEC of finite thickness $\Delta x$.

Then, the energy current across the surface $ \mathcal{A}$  follows from the expression of the energy flux Eq.~\eqref{eq : energy flux} combined with Eq.~\eqref{eq : Phi(x)} as
\begin{equation}
I_E\equiv J_E \mathcal{A} =K\left( \Phi_l - \Phi_r \right).
\label{eq : energy current}
\end{equation}
Once again, we remark that the expression of $I_E$ given in Eq.~\eqref{eq : energy current} takes the form of Fourier's law 
generalized to the composite potential $\Phi$ which also encapsulates the coupling between charge and energy transport. We can finally express $\Phi$ in term of the thermoelectric coefficients and the potentials of the reservoirs as
\begin{equation}
\Phi(x)=\left(\Delta T - \frac{F}{K} I_C \right) \frac{x}{\Delta x}+ T_l + \frac{(\alpha T_l+V_l)^2}{2KR}
\end{equation}
where we introduce
\begin{equation}
F=\frac{\varphi_l+\varphi_r}{2}=\alpha\bar{T}+\bar{V}, \label{eq : F}
\end{equation}
with the mean temperature $\bar{T}=(T_l+T_r)/2$ and electric potential $\bar{V}=(V_l+V_r)/2$.
Then, the second current-force characteristics providing the energy current as function of temperature and voltage differences at the boundaries of the TEC reads
\begin{equation}
I_E=-K\Delta T + F I_C.
\label{eq : global energy current}
\end{equation}
Interestingly, $F$ can be expressed in terms of the charge current by using Eq.~\eqref{eq : global matter current} 
\begin{eqnarray}
F &=& \varphi_l-  \frac{RI_C}{2}\label{eq : Fl}, \\
F &=& \varphi_r +  \frac{RI_C}{2},
\label{eq : Fr}
\end{eqnarray}
implying that Eq.~\eqref{eq : global energy current} is indeed a non linear characteristic equation. As previously mentioned, in the absence of any purely conductive term, that is when $K=0$, the matter and energy currents are strictly proportional. Hence, the parameter $F$ indeed characterizes the free fraction of transported energy.

Eqs.~\eqref{eq : global matter current} and \eqref{eq : global energy current} constitute the integrated current-force characteristics of a TEC where the forces are the temperature and electric potential differences.  

\subsubsection{Temperature and electric potential profile}
\label{sec : profile}

The temperature inside the TEC is the solution of Domenicalli's equation~\cite{Apertet2013} arising from the conservation of the total energy flux $dJ_E/dx=0$ and Eqs.~(\ref{eq : electric flux},\ref{eq : je en fonction de j et nabla T})
\begin{equation}
\frac{d^2T}{dx^2} = - \frac{ J_C^2}{\kappa_J\sigma_T}.
\label{eq : Domenicalli equation}
\end{equation}
The above equation combined with the relations between the second derivatives of $T$ and $V$ following from Eq.~\eqref{eq : linear relation dT and dV} gives the differential equation for the electric potential 
\begin{equation}
\frac{d^2V}{dx^2}= \frac{\alpha J_C^2}{\kappa_J\sigma_T}.
\end{equation}
For our TEC under the boundary conditions of Eqs.~(\ref{leftBC}--\ref{rightBC}), the temperature and electric potential read
\begin{eqnarray}
T &=& T_l +  \frac{x}{\Delta x}\left[ \Delta T-\frac{RI_C^2}{2K}\left( \frac{x}{\Delta x}-1\right) \right] \label{potentialprofile}, \\
V &=& V_l +  \frac{x}{\Delta x}\left[ \Delta V+\frac{\alpha RI_C^2}{2K}\left( \frac{x}{\Delta x}-1\right) \right] \label{temperatureprofile}.
\end{eqnarray}
Let's remarks that, from the temperature gradient
\begin{equation}
\frac{dT}{dx} =\frac{1}{\Delta x} \left[\Delta T - \frac{RI_C^2}{2K}\left( \frac{2x}{\Delta x} - 1 \right) \right] 
\label{eq : nabla T}
\end{equation}
and Eqs.~(\ref{eq : def of phi},\ref{eq : je en fonction de j et nabla T}), we find
\begin{equation}
I_E=-K\Delta T+ \left[\alpha T +V +\frac{RI_C}{2}\left( \frac{2x}{\Delta x} - 1 \right)\right]I_C .
\label{eq : integrated energy flux}
\end{equation}
From energy conservation,
we have exhibited that 
\begin{equation}
	F(x)\equiv \varphi(x) + \frac{RI_C}{2}\left( \frac{2x}{\Delta x} - 1 \right)=\mathrm{const}.
\end{equation}
is a non trivial invariant with respect to the position $x$ within the thermoelectric material. The expressions of $F$ found in Eqs.(\ref{eq : Fl}--\ref{eq : Fr}) are simply boundary values of the invariant $F(x)$ on the left and on the right sides.

\subsection{Levels of description}
\label{sec : non equilibrium conductance matrix}

In Ref.~\cite{Raux2024}, we used different levels of description of thermodynamic devices in view of connecting them. The fundamental level provides a non redundant basis of currents and forces~\cite{Polettini2016}. The level of physical currents and forces, although linearly related, is also of interest: First, it determines all the exchanges with the system's environment; second, it helps to change of basis of fundamental currents. 

In section \ref{SelectionMatrix}, we exhibit a useful form of the selection matrix that relates physical and fundamental currents. This leads us to a specific structure 
of the nonequilibrium conductance matrix at the level of physical currents. In section \ref{subsec : non independent currents for the TEC}, we identify two sets of currents for the description of a TEC: the internal currents crossing the thermoelectric device and the external currents exchanged with the environment. In section~\ref{fundacurrentforce}, we provide two examples of selection matrices allowing to switch from internal (respectively external) physical currents and forces to internal (respectively external) fundamental currents and forces. 
\subsubsection{Structure of the conductance matrices}
\label{SelectionMatrix}

Following the notation of Ref.~\cite{Raux2024}, we denote $\bm i$ the vector of physical currents and $\bm \ell$ the matrix whose lines are the $|\mathscr{L}|$ linearly independent conservation laws relating the $|\mathscr{P}|$ currents in $\bm i$. This writes in matrix form
\begin{equation}
\bm \ell \bm i=\bm 0.
\label{eq : conservation laws}
\end{equation}
Note that if $\bm \ell$ is not full row rank, it should be reduced to a full row rank matrix by removing appropriated raws. The rank-nullity theorem states that the dimension of the kernel of $\bm \ell$ is $\mathrm{dim(\mathrm{ker}(\bm \ell))}=|\mathscr{P}|-|\mathscr{L}| = |\mathscr{I}|$ that is the number of linearly independent currents (i.e., fundamental currents). Eq.~\eqref{eq : conservation laws} thus means that
\begin{equation}
\bm i = \bm S \bm I \quad \text{ with } \quad  \bm \ell \bm S=\bm 0,
\label{eq : selection of independent currents}
\end{equation}
where we have introduced the selection matrix $\bm S$ whose columns constitute a basis of $\mathrm{ker}(\bm \ell)$ and the fundamental currents vector $\bm I$ with $|\mathscr{I}|$ components. 
We show in Appendix \ref{sec : structure of the selection matrix} that the selection matrix can be written as:
\begin{equation}
\bm S=
\begin{bmatrix}
\mathbb{1}_{|\mathscr{I}|}\\
\bm T
\end{bmatrix},
\label{eq : structure of S}
\end{equation}
with $\mathbb{1}_{|\mathscr{I}|}$ the identity matrix of dimension $|\mathscr{I}|$ and $\bm T$ a matrix that only depends on $\bm \ell$. The above form of selection matrix requires that the order of the components of $\bm i$ is chosen such that
\begin{equation}
\bm i =
\begin{pmatrix}
\bm I\\
\bm i_\mathrm{d}
\end{pmatrix}.
\label{eq : splitting i}
\end{equation}
The vector $\bm i_\mathrm{d}$ of the last $|\mathscr{L}|$ components of $\bm i$ are linear functions of $\bm I$. Since the Entropy Production Rate (EPR) $\sigma $ is independent of the level of description, thermodynamics consistency ensures that
\begin{equation}
\sigma=\bm a^T \bm i=\bm A^T\bm I,
\label{eq : EPR dependent = independent}
\end{equation}
i.e., the EPR is a function of the physical force $\bm a$  and its conjugated  current $\bm i$, or similarly of the fundamental force $\bm A$ and current $\bm I$.
Using Eq.\eqref{eq : selection of independent currents} in Eq.~\eqref{eq : EPR dependent = independent} yields
\begin{equation}
\bm A^T=\bm a^T \bm S.
\label{eq : relation fundamental forces physical forces}
\end{equation}
Then, for a current-force relations reading
\begin{align}
\bm i&=\bm g \bm a, 
\label{eq : physical flux force relation} &&\text{(physical level)} \\
\bm I &= \bm G \bm A,
\label{eq : fundamental current-force relation} && \text{(fundamental level)},
\end{align}
the nonequilibrium conductance matrices $\bm g$ and $\bm G$, at respectively the physical and fundamental levels, satisfy
\begin{equation}
\bm g = \bm S \bm G \bm S^T = 
\begin{bmatrix}
\bm G & \bm G \bm T^T\\
\bm T\bm G & \bm T \bm G \bm T^T
\end{bmatrix}.
\label{eq : physical conductance matrix vs fundamental conductance matrix}
\end{equation}
This follows from inserting Eqs.~\eqref{eq : relation fundamental forces physical forces} and \eqref{eq : fundamental current-force relation} in Eq.~\eqref{eq : selection of independent currents}. Therefore, to identify the fundamental conductance matrix knowing the physical conductance matrix in a conveniently chosen basis, it suffices here to read the upper left diagonal sub-matrix of dimension $|\mathscr{I}|$. 
\subsubsection{Physical currents and forces: external, internal}
\label{subsec : non independent currents for the TEC}
\begin{figure}[t]
\centering
\includegraphics[width=0.9\columnwidth]{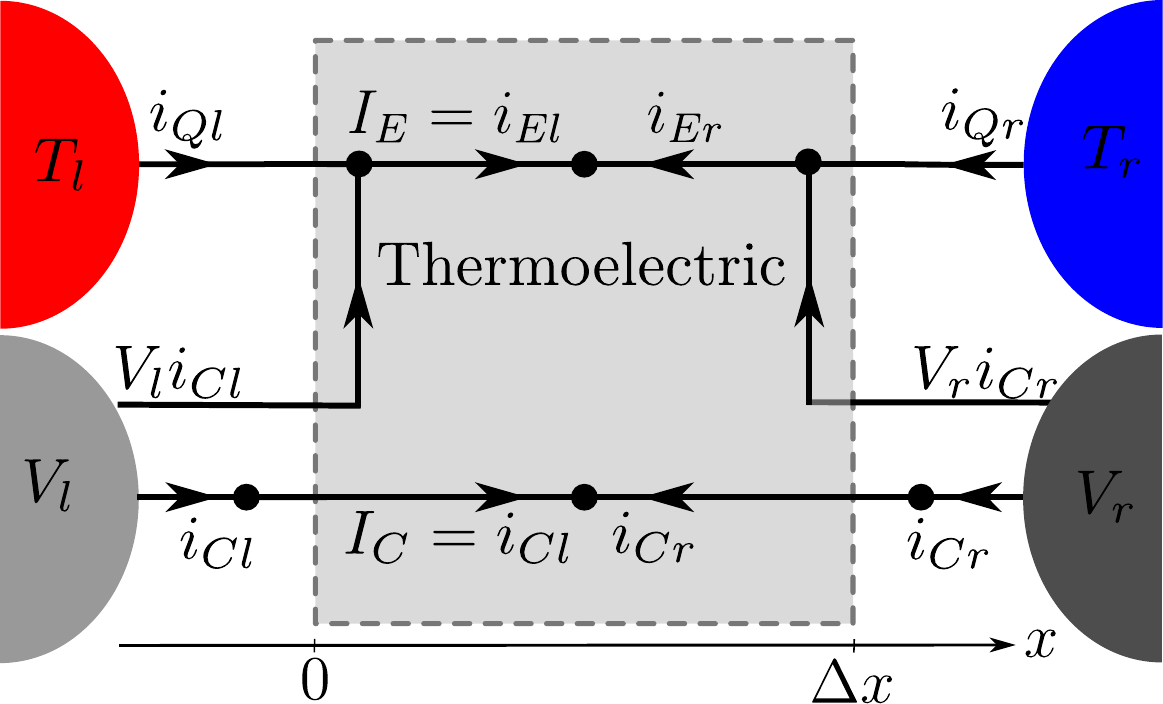}
\caption{The TEC is connected to two thermostats at temperature $T_l$ and $T_r$ with $\Delta T = T_{r}-T_{l} < 0$ and to two metallic leads 
at electric potentials $V_l$ and $V_r$ with $\Delta V=V_r-V_l>0$. At each black dots, the Kirchoff current law can be applied. 
}
\label{fig : schema lois de conservation}
\end{figure}
Our sign convention for currents is summarized on Fig.~\ref{fig : schema lois de conservation}: Physical currents are positive when received by the thermoelectric material in the interval $[0,\Delta x]$. Physical currents entering from the reservoir on the $\chi=l,r$ side (either left or right) are: the (electric) charge current $i_{C\chi}$, the heat current $i_{Q\chi}$ and the energy current $i_{E\chi}$. The electric and energy currents incoming from the left are opposite to those incoming from the right:
\begin{align}
i_{El} =&\,  I_E = - i_{Er} \label{energyconservation}\\
i_{Cl} =& \, \;I_C\; = - i_{Cr}. \label{chargeconservation}
\end{align}
We remark that one can switch of sign convention by considering capitalized currents which are by definition counted as positive when flowing from left to right (in the growing $x$ direction). 
The heat current incoming from the $\chi$ side writes
\begin{eqnarray}
	i_{Q\chi} & =& i_{E\chi}-V_{\chi} i_{C\chi}.
\end{eqnarray}
Energy conservation of Eq.~\eqref{energyconservation} implies that the work current received by the TEC is
\begin{equation}
	i_{W} \equiv -i_{Ql}-i_{Qr} = -I_C \Delta V. \label{workdefinition}
\end{equation}
The conservation laws of physical currents introduced above can be considered at the interface between the system and its environment (grey dashed line in Fig.~\ref{fig : schema lois de conservation}) or across a plane of constant $x\in [0,\Delta x]$. To distinguish them, we call the former \emph{external} physical currents denoted 
\begin{equation}
\bm i_\mathrm{e}^T\equiv
\begin{pmatrix}
i_{Ql} & V_li_{Cl} & i_{Cl} & i_{Qr} & V_r i_{Cr} & i_{Cr} 
\end{pmatrix}, \label{externalcurrents}
\end{equation}
and the latter the \emph{internal} physical currents denoted
\begin{equation}
\bm i_\mathrm{i}^T=
\begin{pmatrix}
i_{El} & i_{Cl} & i_{Er} & i_{Cr} 
\end{pmatrix}. \label{internalcurrents}
\end{equation}
Accordingly, the subscripts ``$\mathrm{e}$'' for external and ``$\mathrm{i}$'' for internal will be used on conjugated and fundamental variables as well. In practice, external currents are useful to make balance with the surrounding leading for instance to the received work above. The internal currents are convenient for the serial association of TECs that we will study in section~\ref{sec : serial association}. Whatever the chosen basis (internal or external), one can always define electric current, heat current, electric power across any transverse plane of the thermoelectric material, etc.
The conservation laws for internal currents (electric charge and energy conservation) write
\begin{equation}
\bm\ell^i\bm i_\mathrm{i}=\bm 0, \; \text{with } \bm \ell^i=
\begin{bmatrix}
1 & 0 & 1 & 0\\
0 & 1 & 0 & 1
\end{bmatrix},
\label{eq : conservation laws internal currents}
\end{equation}
and those for external currents writes
\begin{equation}
\bm \ell^e \bm i_\mathrm{e}=\bm 0, \; \text{with } \bm \ell^e =
\begin{bmatrix}
1 & 1 & 0 & 1 & 1 & 0\\
0 & 0 & 1 & 0 & 0 & 1\\
0 & 1 & -V_l & 0 & 0 & 0\\
0 & 0 & 0 & 0 & 1 & -V_{r}
\end{bmatrix}.
\label{eq : conservation laws external currents}
\end{equation}
Conservation of electric charge and energy is complemented by two additional conservation laws that take into account the proportionality between the electric works $V_\chi i_\chi$ and the electric currents $i_\chi$ on each $\chi=l,r$ side. Finally, we switch between internal and external currents using
\begin{equation}
\bm i_\mathrm{i}=
\bm  P
\bm i_\mathrm{e},\quad \bm i_\mathrm{e}=\bm M \bm i_\mathrm{i},
\label{eq : relation internal physical currents vs external physical currents}
\end{equation}
where 
\begin{equation}
\bm P=
\begin{bmatrix}
1 & 1 & 0 & 0 & 0 & 0\\
0 & 0 & 1 & 0 & 0 & 0 \\
0 & 0 & 0 & 1 & 1 & 0\\
0 & 0 & 0 & 0 & 0 & 1
\end{bmatrix},\quad 
\bm M=
\begin{bmatrix}
1 & -V_l & 0 & 0 \\
0 & V_l & 0 & 0\\
0 & 1 & 0 & 0\\
0 & 0 & 1 &-Vr\\
0 & 0 & 0 & Vr\\
0 & 0 & 0 & 1
\end{bmatrix}.
\end{equation}
We notice that $\bm P \bm M = \mathbb{1}_{4}$, but $\bm M \bm P \neq \mathbb{1}_{6} $ even though $\bm i_\mathrm{e}=\bm M  \bm P \bm i_\mathrm{e} $ is verified for the vector of external currents of Eq.~\eqref{externalcurrents}. We also emphasize that, although $\bm P$ has linearly independent lines and $\bm M$ has linearly independent columns, one should not use their respectively right and left Moore-Penrose pseudo-inverse. This is clear for matrix $\bm M$ that involves the electric potential which creates problems of physical dimensions when using pseudo-inverses. More importantly, and contrarily to what would come from using pseudo-inverse, matrices $\bm P$ and $\bm M$ provides the appropriated forces conjugated to physical currents in the EPR. Indeed, thermodynamic consistency requires that $\sigma $ is independent of the set of variables used to express it
\begin{equation}
\sigma=\bm a_\mathrm{e}^T \bm i_\mathrm{e}=\bm a_\mathrm{i}^T \bm i_\mathrm{i}.
\label{eq : EPR external currents}
\end{equation}
In the stationary state, the EPR is 
\begin{equation}
\sigma=-\sum_{\chi=l,r}\frac{i_{Q\chi}}{T_{\chi}}=-\sum_{\chi=l,r} \frac{i_{E\chi}-V_\chi i_{C\chi}}{T_\chi}.
\end{equation}
This leads to the physical forces conjugated to internal currents
\begin{equation}
\bm a^T_\mathrm{i}=
\begin{pmatrix}
-\frac{1}{T_l} & \frac{V_l}{T_l} & -\frac{1}{T_r} & \frac{V_r}{T_r}
\end{pmatrix}.
\label{eq : physical forces conjugated to the internal currents}
\end{equation}
Using the first relation of Eq.~\eqref{eq : relation internal physical currents vs external physical currents} in Eq.~\eqref{eq : EPR external currents} leads to the physical force conjugated to external currents
\begin{equation}
\bm a_\mathrm{e}^T=\bm a_\mathrm{i}^T\bm P,
\end{equation}
reading explicitly
\begin{equation}
\bm a_\mathrm{e}^T=
\begin{pmatrix}
-\frac{1}{T_l} & -\frac{1}{T_l} & \frac{V_l}{T_l} & -\frac{1}{T_r} & -\frac{1}{T_r} & \frac{V_r}{T_r}
\end{pmatrix}. \label{externalforces}
\end{equation}
Now using the second relation of Eq.~\eqref{eq : relation internal physical currents vs external physical currents} in Eq.~\eqref{eq : EPR external currents}, we obtain a second relation for the forces:
\begin{equation}
\bm a_\mathrm{i}^T=\bm a_\mathrm{e}^T \bm M \label{internalexternalforces}
\end{equation}
which recovers Eq.~\eqref{eq : physical forces conjugated to the internal currents} ensuring the consistency between the description of the TEC using internal or external currents.
\subsubsection{Fundamental currents and forces: external, internal \label{fundacurrentforce}}
In the previous section, we have identified the physical currents and conjugated forces (internal and external). In this section, we provide some possible fundamental currents and forces by choosing selection matrices in the kernel of the matrix of conservation laws. For the simple TEC considered here, we could proceed directly by inspection of the EPR. Indeed one can use the conservation laws directly in this EPR to produce a linear combination of fundamental currents only, with each linear coefficient being the conjugated force. However, we aim here at illustrating the general method that is convenient in more complex situations.

Let's start by choosing some fundamental currents and forces among external physical currents. There are $|\mathscr{P}_\mathrm{e}|=6$ linearly dependent external currents and $|\mathscr{L}_\mathrm{e}|=4$ associated conservation laws. We can thus choose $|\mathscr{I}_\mathrm{e}| = |\mathscr{P}_\mathrm{e}|-|\mathscr{L}_\mathrm{e}|=2$ linearly independent currents by removing $|\mathscr{L}_\mathrm{e}|=4$ currents from $\bm i_\mathrm{e}$ (one per conservation law). 
For instance, we can choose for fundamental currents
\begin{equation}
\bm I_\mathrm{e}=
\begin{pmatrix}
i_{Ql}\\
V_l i_{Cl}
\end{pmatrix}.\label{ChoiceFundExt}
\end{equation}
The selection matrix associated to this choice is:
\begin{equation}
\bm S_\mathrm{e}=
\begin{bmatrix}
\mathbb{1}_2\\
\bm T_\mathrm{e}
\end{bmatrix}, \, \text{ with } \bm T_\mathrm{e}=
\begin{bmatrix}
0 & {1}/{V_l}\\
-1 & {\Delta V}/{V_l}\\
0 & -{V_r}/{V_l}\\
0 & -1/V_{l}
\end{bmatrix}. \label{externalselectionmatrix}
\end{equation}
One can check that $\bm \ell_\mathrm{e}\bm S_\mathrm{e}=\bm 0$. The thermodynamic forces conjugated to the fundamental currents of Eq.~\eqref{ChoiceFundExt} follows from Eq.~\eqref{eq : relation fundamental forces physical forces}
\begin{equation}
\bm A_\mathrm{e}=
\begin{pmatrix}
\frac{1}{T_r}-\frac{1}{T_l}\\
-\frac{1}{T_r}\frac{\Delta V}{V_l}
\end{pmatrix}.
\end{equation}

Let's continue with choosing some fundamental currents and forces among internal physical currents. There are $|\mathscr{P}_\mathrm{i}|=4$ linearly dependent internal currents and $|\mathscr{L}_\mathrm{i}|=2$ associated conservation laws. As above, we choose $|\mathscr{I}_\mathrm{i}|  = |\mathscr{P}_\mathrm{i}|-|\mathscr{L}_\mathrm{i}|=2$ linearly independent currents by removing $|\mathscr{L}_\mathrm{i}|=2$ currents from $\bm i_\mathrm{i}$ (one per conservation law). For instance, we can choose for fundamental currents
\begin{equation}
\bm I_\mathrm{i}=
\begin{pmatrix}
i_{El}\\
i_{Cl}
\end{pmatrix} = \begin{pmatrix}
I_{E}\\
I_C
\end{pmatrix}. \label{ChoiceFundInt}
\end{equation}
The selection matrix associated to this choice is:
\begin{equation}
\bm S_\mathrm{i}=
\begin{bmatrix}
\mathbb{1}_2\\
-\mathbb{1}_2
\end{bmatrix}. \label{internalselectionmatrix}
\end{equation}
We notice that $\bm S_\mathrm{i}$ has the structure given in Eq.~\eqref{eq : structure of S} with $\bm T_\mathrm{i}=-\mathbb{1}_2$. Here again, one can check that $\bm \ell_\mathrm{i}\bm S_\mathrm{i}=\bm 0$. The thermodynamic forces conjugated to the fundamental currents of Eq.~\eqref{ChoiceFundInt} follows 
\begin{equation}
\bm A_\mathrm{i} =
\begin{pmatrix}
\frac{1}{T_r}-\frac{1}{T_l}\\
\frac{V_l}{T_l}-\frac{V_r}{T_r}
\end{pmatrix} \equiv \begin{pmatrix}
A_{E}\\
A_{C}
\end{pmatrix} ,
\end{equation}
where we have introduced for later convenience the fundamental force $A_{E}$ (respectively $A_{C}$) conjugated to energy (respectively electric) current. By construction, the EPR of the TEC is
\begin{equation}
    \sigma = \bm a_\mathrm{e}^T \bm i_\mathrm{e}=\bm A_\mathrm{e}^T \bm I_\mathrm{e}= \bm a_\mathrm{i}^T \bm i_\mathrm{i}=\bm A_\mathrm{i}^T  \bm I_\mathrm{i}.
    \label{eq : entropy production}
\end{equation}

\subsection{Conductance matrices for physical and fundamental variables}
\label{CondMatrix}
In this section, restarting from Ioffe's current-force characteristics for a CPM, we introduce the conductance matrix of  a TEC at fundamental level. We then go on with the derivation of the conductance matrix at all level of description. In particular, we show how to derive the conductance matrix at fundamental scale coupling heat and electric current. 

According to Eqs.~(\ref{eq : global matter current}--\ref{eq : global energy current}), the current-force characteristic of the TEC can be arranged in matrix form as
\begin{equation}
\bm I_\mathrm{i} =\begin{pmatrix}
I_{E}\\
I_C
\end{pmatrix} = 
-\frac{1}{R}
\begin{bmatrix}
 \alpha F + KR  &   F \\
 \alpha & 1
\end{bmatrix}
\begin{pmatrix}
\Delta T \\
\Delta V
\end{pmatrix}.
\label{eq : Onsager delta T delta mu}
\end{equation}
where appears a non-symmetric matrix synonymous of non-conjugated currents and forces. We recover a symmetric matrix by switching to conjugated forces via 
\begin{equation} 
 \begin{pmatrix}
\Delta T \\
\Delta V
\end{pmatrix} =-\frac{T_rT_l}{\bar T} 
\begin{bmatrix}
\bar{T} & 0 \\
\bar{V}  & {1}
\end{bmatrix}
\bm A_\mathrm{i}
\label{eq : force change of basis}
\end{equation}
yielding the current-force relation 
\begin{equation}
\bm I_\mathrm{i}
=\bm G_\mathrm{i}
\bm A_\mathrm{i}
\label{eq : fundamental current-force relation for internal currents}
\end{equation}
with
\begin{equation}
\bm G_\mathrm{i}
=\frac{T_lT_r}{R\bar{T}}
\begin{bmatrix}
F^2 + KR \bar{T} & F \\
F & 1
\end{bmatrix}.
\label{eq : global flux force relation conjugated}
\end{equation}
This conductance matrix is symmetric and positive definite, as required for a TEC arbitrarily far from equilibrium. 
It is also force dependent, here via the $\bar T$ and $\bar V $ appearing in $F$.
The conductance matrix $\bm G_\mathrm{i}$ provides not only the non-linear current-force characteristics, but also constitutes a modelling choice of the coupling between energy and electric currents. As such, it represents an extension of the constant property model for thermoelectric~\cite{Goupil2011, Apertet2012} although no microscopic modeling has been required.

In Appendix \ref{sec : non unicity of the conductance matrices}, we show that the above conductance matrix is in principle not unique. For given mean currents and conjugated forces, making a choice of conductance matrix amounts to constrain the currents covariance~\cite{Horowitz2020,Vroylandt2018,Vroylandt2019}. 
However, the conductance of Eq.~\eqref{eq : global flux force relation conjugated} is the simplest: any other form involves the supplementary definition of a bivariate function of fundamental forces.

We end this section by providing the conductance matrix at the physical level (external and internal) from the one at the fundamental level. First, we apply Eq.~\eqref{eq : physical conductance matrix vs fundamental conductance matrix} using the selection matrix of Eq.~\eqref{internalselectionmatrix} leading to the conductance for internal physical currents
\begin{equation}
\bm g_\mathrm{i}=\bm S_\mathrm{i}\bm G_\mathrm{i}\bm S_\mathrm{i}^T=
\begin{bmatrix}
\bm G_\mathrm{i} & -\bm G_\mathrm{i}\\
-\bm G_\mathrm{i} & \bm G_\mathrm{i}
\end{bmatrix}, \, \text{ for } \bm i_\mathrm{i} = g_\mathrm{i} \bm a_\mathrm{i}. \label{eq : G physical level}
\end{equation}
As expected this matrix is symmetric, although it is now semi-definite positive only. 
Second, we use Eqs.~\eqref{eq : relation internal physical currents vs external physical currents} and \eqref{internalexternalforces} to write
\begin{equation}
	\bm i_\mathrm{e} = \bm M \bm i_\mathrm{i} = \bm M \bm g_\mathrm{i} \bm a_\mathrm{i} = \bm M \bm g_\mathrm{i} \bm M^{T} \bm a_\mathrm{e}
\end{equation}
from which we read the conductance $\bm g_\mathrm{e} = \bm M \bm g_\mathrm{i} \bm M^T$ for external physical currents. We give this conductance matrix in Table \ref{eq: conductance matrix physical external} since it can be used conveniently to extract the conductance matrix at the fundamental level for any choice of fundamental currents.
\begin{table*}
\begin{equation}
\bm g_\mathrm{e} = \frac{T_lT_r}{R\bar{T}}
\begin{bmatrix}
 (F-V_l)^2+ K R\bar{T} & V_l (F-V_l) & F-V_l & -\left((F-V_l) (F-V_r)+KR\bar{T}\right)  &V_r (V_l-F) & V_l-F\\
V_l (F-V_l) & V_l^2 & V_l & V_l (V_r-F) & -V_l V_r & -V_l \\
 F-V_l & V_l & 1 & V_r-F & -V_r & -1\\
 -\left((F-V_l) (F-V_r)+KR\bar{T}\right)  & V_l (V_r-F) & V_r-F &  (F-V_r)^2+ K R\bar{T} & V_r (F-V_r) & F-V_r\\
V_r (V_l-F) & -V_l V_r & -V_r & V_r (F-V_r) & V_r^2 & V_r\\
 V_l-F & -V_l & -1 & F-V_r & V_r & 1
\end{bmatrix}.
\nonumber
\end{equation}
\caption{Conductance matrix for physical currents and forces given in Eqs.~\eqref{externalcurrents} and \eqref{externalforces}. \label{eq: conductance matrix physical external}}
\end{table*}
To do so, one simply select $\bm g$'s components for the lines and columns associated to the chosen currents and forces. We remark that conjugated forces must still be determined using the appropriate selection matrix for chosen currents. For instance, one has the following current-force relation at the fundamental level  
\begin{equation}
\begin{pmatrix}
i_{Ql}\\
i_{Cl}
\end{pmatrix}
=\begin{bmatrix}
	g_{11} & g_{13}\\
	g_{31} & g_{33}
	\end{bmatrix}
\begin{pmatrix}
\frac{1}{T_r}-\frac{1}{T_l}\\
-\frac{\Delta V}{T_r}
\end{pmatrix}.
\end{equation}
We use table \ref{eq: conductance matrix physical external} to read directly the conductance matrix 
\begin{equation}
	\begin{bmatrix}
	g_{11} & g_{13}\\
	g_{31} & g_{33}
	\end{bmatrix}=\frac{T_lT_r}{R\bar{T}}
\begin{bmatrix}
KR\bar{T}+(F-V_l)^2 & F-V_l\\
F-V_l & 1
\end{bmatrix}.
\end{equation} 
The degree of coupling $\xi \in [-1,1] $ between the heat and charge currents entering from the left of the TEC, respectively $i_{Ql}$ and $i_{Cl}$, follows 
\begin{equation}
\xi= \frac{g_{13}}{\sqrt{g_{11}g_{33}}} = \frac{F-V_l}{\sqrt{KR\bar{T}+(F-V_l)^2}} .
\end{equation}
It generalizes far from equilibrium the notion of degree of coupling that has a fundamental role in conversion processes \cite{Kedem1965, Polettini2015, Vroylandt2018}.

\section{Association of TECs in stationary
non-equilibrium}
\label{Association}

Based on the description of TEC established in the first part, we study in this section the serial and parallel association of two TECs. Superscripts $(m)$ for $m=1,2$ are added to any physical variable or parameter of the first part to indicate which TEC it characterises in this second part.

\subsection{Serial association}
\label{sec : serial association}
\begin{figure*}[t!]
    \centering
    \includegraphics[width=1.7\columnwidth]{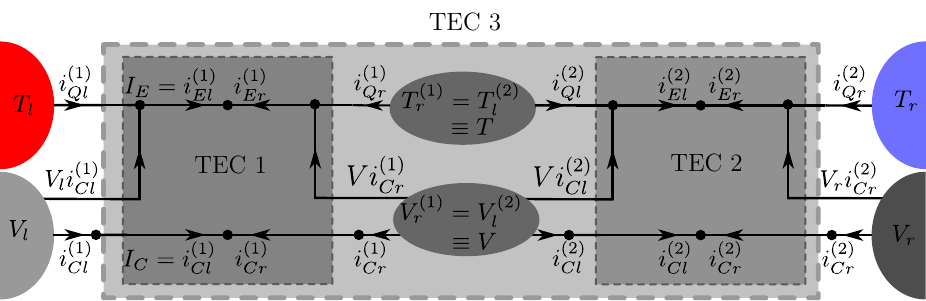}
    \caption{Sketch of TEC $3$ obtained as the serial association of TEC 1 and 2. 
    TEC 1 is connected to a thermostat at temperature $T_l$ on its left and to the thermal pin of TEC 2 at temperature $T$ on its right. Similarly, it is connected to a metallic lead on its left which set electrical potential to $V_l$ and to the electric pin of TEC 2 at voltage $V$ on its right. The situation for TEC 2 is the left-right symmetric. 
    TEC 3 is connected to two thermostats at temperatures $T_l$ and $T_r$ with $\Delta T^{(3)}=T_r-T_l<0$ and to two metallic leads at electrical potentials $V_l$ and $V_r$ with $\Delta V^{(3)}=V_l-V_r>0$.
    The sum of currents (incoming arrows) at each black bullet is zero.\label{fig : composite TEC}}
\end{figure*}

Fig.~\ref{fig : composite TEC} illustrates the serial association of TECs 1 and 2 leading to TEC 3. It defines the boundary conditions for TEC 3 and emphasizes the conservation of energy and electric currents at the interface of TECs 1 and 2:
\begin{align}
I_E^{(1)}&=I_E^{(2)}=I_E,
\label{eq : energy conservation at interface}\\
I_C^{(1)}&=I_C^{(2)}=I_C
\label{eq : charge conservation at interface},
\end{align} 
where $I_E^{(m)}$ and $I_C^{(m)}$ are defined as in Eqs.~(\ref{energyconservation}--\ref{chargeconservation}) as the currents (positive when) entering into device $m=1,2$ from the left. Given charge and energy conservation, we drop the superscript $(m)$ for these currents to shorten notation.
Since dissipation only occurs inside the elements no dissipation occurs at the interface between TECs 1 and 2 which implies the following potential continuity equations: 
\begin{align}
T_r^{(1)}&=T_l^{(2)}\equiv T,
\label{eq : temperature equality at interface}\\
V_r^{(1)}&=V_l^{(2)}\equiv V,
\label{eq : electrical potential at interface}
\end{align}
where we have introduced $T$ the local temperature and $V$ the electric potential at the interface to shorten notation. It is also convenient to remove superscripts for the outter boundary condition of TEC 3 by denoting $T_{l}=T_{l}^{(1)}=T_{l}^{(3)}$ and $T_{r}=T_{r}^{(2)}=T_{r}^{(3)}$, and similarly for the electric potential.

We start by the determination of $T$ and $V$ in section \ref{subsec : internal local potentials}. Then, we get the equivalent conductance matrix describing the current-force characteristics of TEC 3 in section \ref{sec : res add} where we use the law of resistance matrix addition~\cite{Raux2024}. Discussion follows in section \ref{subsec : discussion}: We compare this approach with an effective model (introduced in Appendix \ref{effective model}) with current dependent thermoelectric coefficients and leading to the same current-force characteristics. Finally, we discuss the physics of the serial association of two TECs where Peltier effect due to differences in Seebeck coefficients appears clearly.

\subsubsection{Internal degrees of freedom: local potentials at the interface}
\label{subsec : internal local potentials}
\paragraph*{Determination of $T$} Combining the energy currents Eq.~\eqref{eq : global energy current} with the conservation of energy at the interface Eq.~\eqref{eq : energy conservation at interface}, we obtain
\begin{equation}
    \left(K^{(1)} + K^{(2)}\right)T - \left(K^{(1)}T_l + K^{(2)}T_r\right)=\left(F^{(1)}-F^{(2)}\right)I_C
    \label{eq : T (1)}
\end{equation}
Using the definition of $F^{(m)}$ for $m=1,2$ introduced in Eq.~\eqref{eq : F}, we simplify $F^{(1)}-F^{(2)}$ as 
\begin{equation}
    F^{(1)}-F^{(2)}=\alpha^{(1)}\bar{T}^{(1)}-\alpha^{(2)}\bar{T}^{(2)}-\Delta V^{(3)}.
    \label{eq : F1-F2}
\end{equation}
with the average temperature and the average electric potential of TEC $m$ reading respectively 
\begin{eqnarray}
	\bar{T}^{(m)}&=& (T_l^{(m)}+T_r^{(m)})/2, \\
 	\bar{V}^{(m)}&=&(V_l^{(m)}+V_r^{(m)})/2.
\end{eqnarray}
Then, using the local potential continuity at the interface and the expression of the electric current Eq.~\eqref{eq : global matter current}, we express $\Delta V^{(3)}$ as
\begin{equation}
    \Delta V^{(3)}=-\sum_{m=1}^2 \left(R^{(m)}I_C + \alpha^{(m)}\Delta T^{(m)}\right).
    \label{eq : delta V3}
\end{equation}
Inserting this last equation in Eq.~\eqref{eq : F1-F2} yields 
\begin{equation}
    F^{(1)}-F^{(2)}=-\delta_\alpha T + \frac{R^{(1)}+R^{(2)}}{2}I_C
\end{equation}
where $\delta_\alpha = \alpha^{(2)}-\alpha^{(1)}$. Finally, by inserting this last relation in Eq.~\eqref{eq : T (1)} and solving for $T$ yields
\begin{equation}
T=\displaystyle \frac{K^{(1)}T_l+K^{(2)}T_r+\frac{R^{(1)}+R^{(2)}}{2}I_C^2}{K^{(1)}+K^{(2)}+\delta_\alpha I_C}.
\label{eq : T}
\end{equation}
Requiring $T>0$ leads to two inequalities according to the sign of $\delta_\alpha$:
\begin{align}
    I_C&>-\frac{K^{(1)}+K^{(2)}}{\delta_\alpha}\;\mathrm{if}\;\delta_\alpha>0,\label{eq : forbidden currents dalpha>0}\\
    I_C&<-\frac{K^{(1)}+K^{(2)}}{\delta_\alpha}\; \mathrm{if}\; \delta_\alpha<0.\label{eq : forbidden currents dalpha<0}
\end{align}

\paragraph*{Determination of $V$} The interface electric potential $V$ follows from using Eq.~\eqref{eq : global matter current} in Eq.~\eqref{eq : charge conservation at interface} since $\Delta V^{(1)}=V-V_l$ and $\Delta V^{(2)}=V_r-V$,
leading to
\begin{equation}
V=R_\parallel\left( \frac{V_l}{R^{(1)}} + \frac{V_r}{R^{(2)}} + \frac{\alpha^{(2)}\Delta T^{(2)}}{R^{(2)}} - \frac{\alpha^{(1)}\Delta T^{(1)}}{R^{(1)}} \right)
\label{eq : V}
\end{equation}
where $R_\parallel=R^{(1)}R^{(2)}/(R^{(1)}+R^{(2)})$. We emphasize that it depends on $I_C$ via $T$ in $\Delta T^{(1)} = T-T_{l}$ and $\Delta T^{(2)} = T_{r}-T$. 

Eqs.~(\ref{eq : T}) and (\ref{eq : V}) show that devices $1$ and $2$ have local potential differences at their boundaries that depends on the electric current even though device $3$ is in Dirichlet boundary conditions. 
\subsubsection{Equivalent resistance matrix} \label{sec : res add}
In Appendix \ref{subsec : conservation laws} we illustrate the general approach of Ref.~\cite{Raux2024} to exhibit the conservation laws of TEC 3 resulting from the serial association of TECs 1 and 2. As expected, we recover the conservation of electric and energy currents. Given the similarity between the devices, the law of resistance matrix addition applies rather trivially for the serial association of TECs 1 and 2, see Appendix~\ref{conductance matrix dimension matching}:
\begin{equation}
    \bm R^{(3)}=\bm R^{(1)} + \bm R^{(2)},
    \label{eq : R1+R2}
\end{equation}
associated to the following force-current characteristics
\begin{equation}
    \begin{pmatrix}
    \frac{1}{T_r}-\frac{1}{T_l}\\
    \frac{V_l}{T_l}-\frac{V_r}{T_r}
    \end{pmatrix}=\bm R^{(3)}
    \begin{pmatrix}
    I_E\\
    I_C
    \end{pmatrix}.
    \label{eq : force current device 3}
\end{equation}
The matrix $\bm R^{(3)}$ reads explicitly in terms of the properties of two TECs
\begin{equation}
\bm R^{(3)}=
\sum_{m=1}^2
\begin{bmatrix}
\frac{1}{K^{(m)}T_l^{(m)}T_r^{(m)}} & -\frac{F^{(m)}}{K^{(m)}T_l^{(m)}T_r^{(m)}}\\
-\frac{F^{(m)}}{K^{(m)}T_l^{(m)}T_r^{(m)}} & \frac{K^{(m)}R^{(m)}\bar{T}^{(m)}+F^{(m)2}}{K^{(m)}T_l^{(m)}T_r^{(m)}}
\end{bmatrix}.
\label{eq : R3 explicit}
\end{equation}
It is positive definite, since each $\bm R^{(m)}$ is positive definite for $m=1,2$. 

\subsubsection{Discussion}
\label{subsec : discussion}

\paragraph*{Equivalence between resistance matrices}

As discussed in Appendix~\ref{sec : non unicity of the conductance matrices}, infinitely many resistance matrices are associated to the same force--current characterics [here Eqs.~(\ref{eq : force current device 3}--\ref{eq : R3 explicit})]. To illustrate this point, we provide in Appendix~\ref{effective model} the resistance matrix of an effective model with identical characteristic equations. If one focuses only on the mean currents, the representative of the resistance matrices of the main text or appendix are equivalent. However, the degree of coupling between energy and electric currents depends on this choice of representative with important consequences. For instance, the degree of coupling constrains the conversion or transduction efficiency~\cite{Vroylandt2018, Wachtel2022}. In the same idea, the nonequilibrium conductance or resistance matrix bounds the quadratic fluctuations of currents as shown in Ref.~\cite{Vroylandt2019}. Therefore, when modeling subsystems with conductance or resistance matrices, we expect that only our theory of thermodynamic circuits predicts the degree of coupling or provides a bound for currents quadratic fluctuations resulting from the equivalent resistance (or conductance) matrix. In other words, there is more in conductance matrices than in a current-force characteristics: they includes subtle details associated to current fluctuations that only the microscopic dynamics can describe in full details.

\paragraph*{Homogeneous TEC} 

In the limit in which TEC 1 and 2 are identical, i.e., for homogeneous thermoelectric coefficients $\alpha^{(1)}=\alpha^{(2)}=\alpha, K^{(1)}=K^{(2)}=K$ and $R^{(1)}=R^{(2)}=R$ and for TECs of same thickness $\Delta x^{(1)} = \Delta x^{(2)}$, we expect to find the characteristic equation of a single TEC of double thickness and thermoelectric coefficients
\begin{align}
    K^{(3)}&=K/2,& 
    R^{(3)}&=2R,&     \alpha^{(3)}&=\alpha.
\end{align}
This is indeed the result of the serial association of a thermal or an electric resistor. The effective model of Appendix~\ref{effective model} has such thermoelectric coefficients in the homogeneous case. From the equivalence of the two models, the characteristic equations obtained from the law of resistance matrix addition (section \ref{sec : res add}) is compatible with this result.

\paragraph*{Non linear voltage-current characteristics}
\begin{figure*}
\centering
\includegraphics[width=\textwidth]{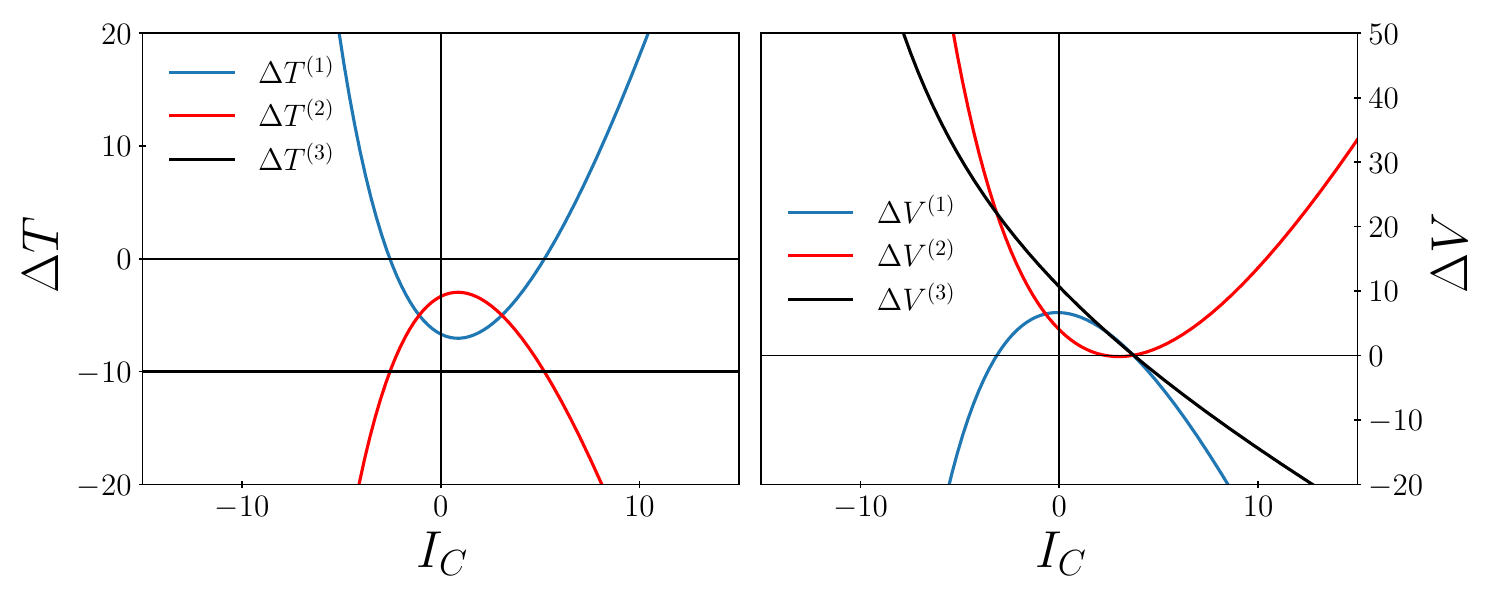}
\caption{Temperatures and voltages differences at the boundary of each TEC as function of the electric current $I_{C}$ at fixed $\Delta T^{(3)}<0$. TEC 1 and 2 are in serial association resulting in TEC 3.
(Left panel)  
Temperature differences: $\Delta T^{(1)}$ (blue solid line), $\Delta T^{(2)}$ (red solid line) and $\Delta T^{(3)}$ (black solid line). 
(Right panel) 
Voltage differences: $\Delta V^{(1)}$ (blue solid line), $\Delta V^{(2)}$ (red solid line) and $\Delta V^{(3)}$ (black solid line).
For both panels, parameters are: $K^{(1)}=1, K^{(2)}=2,R^{(1)}=1,R^{(2)}=2, \alpha^{(1)}=1,\alpha^{(2)}=1.2$, $T_l=20,T_r=10$ and $V_l=10$. 
We use an ad hoc unit system: $T_{r}/10=1$ set the temperature unit, Boltzmann constant $k_{B} = 1$ defines the energy unit, thermal conductivity defines the time unit $k_{B}/K^{(1)} = 1$, Seebeck coefficient and electric resistance of TEC 1 define the unit of electric current via $\alpha^{(1)}T_{r}/10R^{(1)} = 1$ meaning that $\alpha^{(1)}T_{r}/10$ is a suitable unit of electric voltage and $R^{(1)}$ of electric resistance.
\label{fig : potentitels au milieu}}
\end{figure*}
Fig.~\ref{fig : potentitels au milieu} shows the temperature differences and voltages applied to the devices 1, 2 and 3. From the effective model for TEC $3$, we can express the voltage-current characteristics  
\begin{equation}
\Delta V^{(3)}=-\alpha^{(3)}\Delta T^{(3)} - R^{(3)}I_C 
\end{equation}
in which $\alpha^{(3)}$ and $R^{(3)}$ are functions of the electric current $I_C$ [see Eqs.\eqref{eq : alpha3} and \eqref{eq : R3}], except in the homogeneous case. Hence, the voltage-current characteristics of TEC 3 is in general non linear under fixed temperature difference $\Delta T^{(3)}$. Since a single TEC has constant thermoelectric coefficient, it has a linear voltage-current characteristics under fixed temperature difference $\Delta T^{(m)}$ for $m=1,2$. Under fixed $\Delta T^{(m)}$, a single TEC is thus analogous to a Thévenin voltage source (ideal voltage source serially connected to a resistor). The serial association of two Thévenin voltage sources would lead to a linear equivalent voltage-current characteristics in the abscence of coupling with energy. We thus interpret the emergeant non linearity as the result of the coupling between energy and charge currents. 

To investigate this coupling, we open device $3$ to take a look at the boundary conditions applied to devices $1$ and $2$ within the serial association. The left panel of Fig.~\ref{fig : potentitels au milieu} shows the temperature differences on device $m=1,2,3$ as a function of $I_C$. By assumption, $\Delta T^{(3)}$ is constant. To accomodate this constraint, the temperature differences applied to devices 1 and 2 are non linear functions of $I_C$. The right panel of Fig.~\ref{fig : potentitels au milieu} shows that the voltage-current characteristics of devices $m=1,2,3$ are all non linear functions of $I_C$, as aforementioned for devices $3$. For devices $1$ and $2$, the non-linearity of $\Delta V^{(1)}$ and $\Delta V^{(2)}$  are due to the $I_C$ dependent $\Delta T^{(1)}$ and $\Delta T^{(2)}$. 
Devices 1 and 2 are thus no longer analogous to Thévenin voltage sources. 

Therefore, the interplay between the energy/charge coupling and non ideal boundary conditions, leads to complex characteristics equations for TECs 1, 2 and 3. These more complex behaviors, due in the end to inhomogenous thermoelectric coefficients, also appear in the spatial profiles of the potentials (temperature, electric potential, composite potentials), as we investigate below.
\paragraph*{Peltier effect at the interface} 
\label{DiscussionPeltier}
\begin{figure*}
\centering
\includegraphics[width=\textwidth]{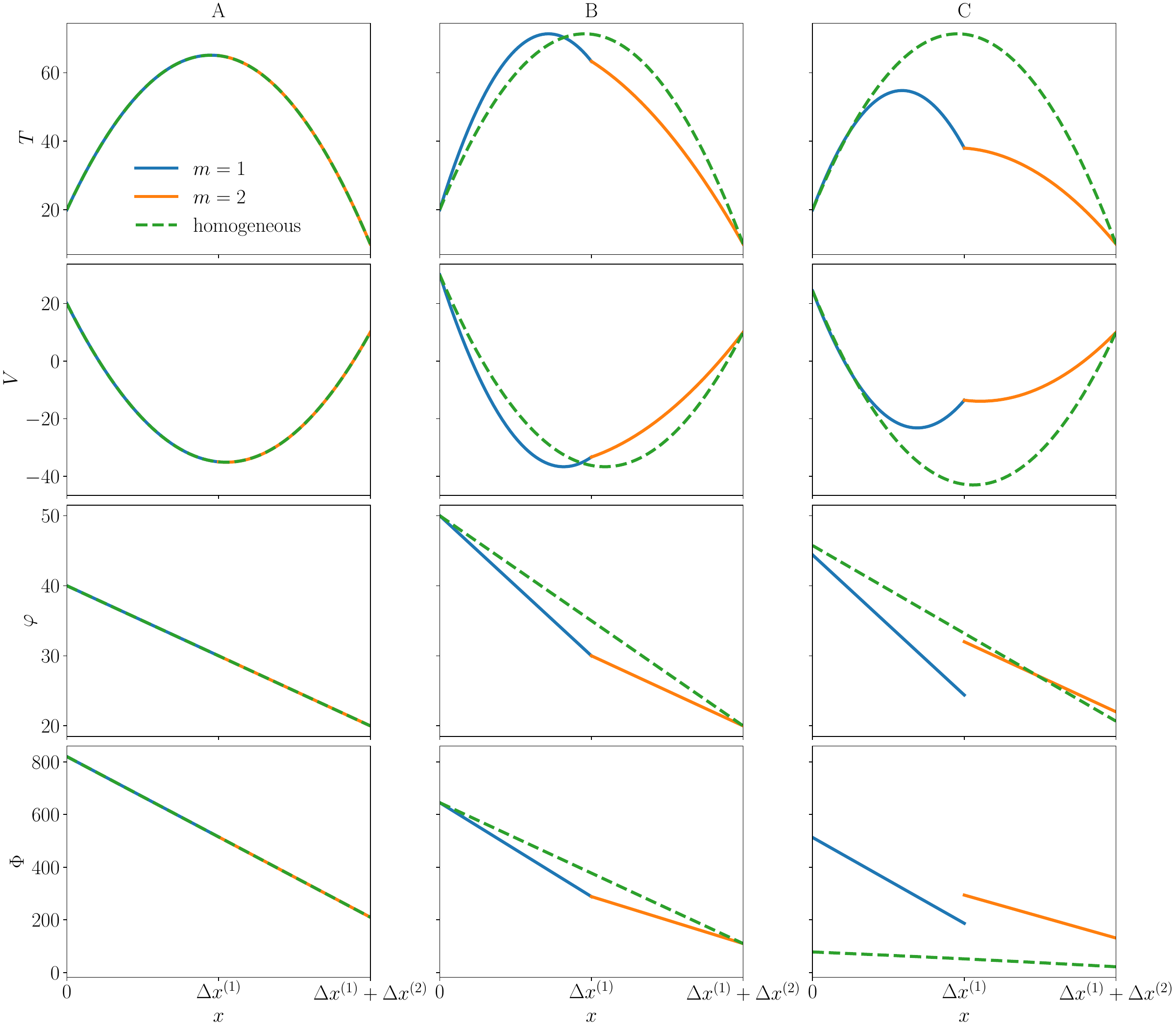}
\caption{Potential profiles $T$, $V$, $\varphi$ and $\Phi$ along the $x$ direction for TEC 1 (blue line), TEC 2 (orange line) and for a TEC with homogenous thermoelectric coefficients $R^{(1)}+R^{(2)}, K^{(1)}K^{(2)}/(K^{(1)}+K^{(2)})$ (green dashed line). Those TECs are respectively of length $\Delta x^{(1)}$, $\Delta x^{(2)}$ and $\Delta x^{(1)}+\Delta x^{(2)}$. Boundary conditions are shown in Fig.~\ref{fig : composite TEC}. The legend  in the upper left panel applies to all panels. The first line of graphs corresponds to the temperature $T(x)$, the second to the electric potential $V(x)$, the third to the composite potential $\varphi(x)$ and the fourth to the composite potential $\Phi(x)$. In each column, the set of thermoelectric coefficients for both TECs are fixed to particular values. Column $\mathrm{A}$ corresponds to TEC 1 and 2 with equal thermoelectric coefficients: $K^{(1)}=K^{(2)}=1, R^{(1)}=R^{(2)}=1,\alpha^{(1)}=\alpha^{(2)}=1$. Column $\mathrm{B}$ corresponds to the weaker condition ensuring composite potential continuity with $K^{(1)}=1,K^{(2)}=2, R^{(1)}=2,R^{(2)}=1,\alpha^{(1)}=\alpha^{(2)}=1$. Column $\mathrm{C}$ corresponds to TEC 1 and 2 with different thermoelectric coefficients $K^{(1)}=1,K^{(2)}=2, R^{(1)}=2,R^{(2)}=1,\alpha^{(1)}=1,\alpha^{(2)}=1.2$. We also fix the boundary conditions $T_l=20,T_r=10,V_r=10, I_C=5$. Same unit system as in Fig.~\ref{fig : potentitels au milieu}.
\label{fig : profiles}}
\end{figure*}
We show on Fig.~\ref{fig : profiles} the spacial profiles of the potentials $T$, $V$, $\varphi$ and $\Phi$ for TEC 1 and 2 in serial association. We consider three different sets of parameters as described below.

In the case of an homogeneous system (column A),  
the serial association of the two TEC is invisible: TEC 3 behaves as it were a single device as studied in section \ref{DefG}. The profiles of the composite potentials are linear as expected from Eqs.~(\ref{eq : phi (1)}--\ref{eq : Phi(x)}). 
The similarity between the quadratic profiles for $T(x)$ and $V(x)$ comes from $\varphi = \alpha T + V$ leading to an affine relation between their gradients (with linear coefficient $-\alpha$) due to Eq.~\eqref{eq : electric flux} and the conservation of electric current deriving from $\varphi$.
We can usefully 
comment the expression linking energy, heat and electric currents
\begin{align}
I_Q(x)=I_E-V(x)I_C=\left(-K \frac{d\Phi}{dx}+\frac{V}{R} \frac{d\varphi}{dx}I_C\right)\Delta x \label{eq : Gibbs-Duhem}.
\end{align}
Although not plotted, we can immediately see that the profile of $I_Q(x)$ is simply given by an affine function of $V(x)$. This naturally leads to the result $dI_Q=-dV(x)I_C$, which translates the conversion of heat into work on a local scale. Note that any non-derivability or discontinuity in $V(x)$ immediately results in a modification of the system's heat balance and heat-to-work conversion. 

The situation of non-derivability of the potentials appears in the case (column B) where the thermal and electrical conductivities differ between TEC 1 and 2, while maintaining the same Seebeck coefficient. In this case, we observe a slope break without discontinuity between the $\varphi$ and $\Phi$ potentials. Indeed, each slope corresponds to the current conductivity ratio that switches value at the interface. Given the conservation of electric (resp. energy) current the slope break of $\varphi$ (resp. $\Phi$) increases with the discrepancy of the electric (resp. thermal) conductivity. This slope breaks also modifies the balance given by Eq.~\eqref{eq : Gibbs-Duhem} as aforementioned. Finally, the temperature and electric potential profiles still share great similarity for the same reason as in the homogeneous case, although now they are quadratic in each TEC separately.  

In the case (column C) where there is in addition a difference between the values of the Seebeck coefficients between TEC 1 and 2, the consequence is much greater. Knowing that the Seebeck coefficient is a measure of entropy per carrier $S_N=e\alpha$ with $e$ the charge per carrier, it follows that any change in the value of the Seebeck coefficient translates into a modification of the energy balance (since heat current depends on the entropy per carrier). This situation is well known in thermoelectricity, as it is nothing less than the manifestation of the Peltier effect at the interface between TEC 1 and 2. In this case, the Peltier contribution to the total heat current $I_{Q} = \mathcal{A} J_{Q}$ undergoes a discontinuity $\delta_\alpha T I_{C}$. As a result, the energy and electric currents do not derive any more from the potentials $\varphi$ and $\Phi$ that are discontinuous at the interface, such that
\begin{eqnarray}
    \varphi_l^{(2)}-\varphi_r^{(1)}&=&\delta_\alpha T, \\
    \Phi_l^{(2)} - \Phi_r^{(1)} &=& \frac{{\varphi_r^{(1)}}^{2}}{2}\left( \frac{1}{K^{(2)}R^{(2)}} - \frac{1}{K^{(1)}R^{(1)}} \right) \nonumber \\
    && +
    \frac{\delta_\alpha T \left( \delta_\alpha T + \varphi_r^{(1)} \right)}{2K^{(1)}R^{(1)}}.
\end{eqnarray}
On Fig.~\ref{fig : profiles}, these discontinuities are positive, resulting in an increase in the values of the $\varphi$ and $\Phi$ potentials at the interface, while their slopes remain identical to those in the previous case. We recover the potential continuity if, and only if, $\delta_\alpha=0$ and $K^{(1)}R^{(1)}=K^{(2)}R^{(2)}$. 

We conclude from the above analysis that the serial association of two TECs satisfying separately the CPM opens a variety of internal degrees of freedom (e.g., potential profiles, current dependent thermoelectric coefficients) that introduce sophisticated behaviors not observed in the CPM, while keeping the degree of complexity reasonably low. This analysis also confirms that the serial association of two identical TECs is totally equivalent to a unique TEC of double length, as expected.

\subsection{Parallel association}
\label{sec : parallel association}
\begin{figure*}
\centering
\includegraphics[width=0.8\textwidth]{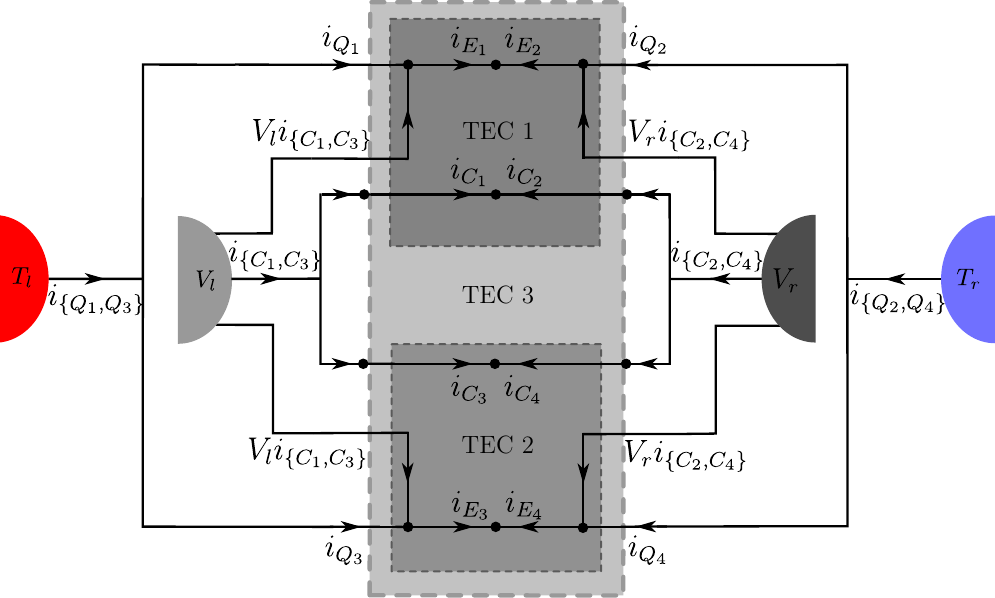}
\caption{Sketch of TEC $3$ obtained as the parallel association of TEC 1 and 2. TEC 1, 2 and 3 are connected to two thermostats at temperatures $T_l$ and $T_r$ with $\Delta T^{(3)}=T_r-T_l<0$ and to two metallic leads at electrical potentials $V_l$ and $V_r$ with $\Delta V^{(3)}=V_l-V_r>0$. \label{fig : parallel TEC}.\label{fig : parallel}}
\end{figure*}
We now turn to the study of the parallel association of two TECs. The equivalent current-force characteristics described by the equivalent conductance matrix is obtained in section~\ref{subsec : equivalent conductance matrix parallel} from the law of conductance matrix addition. We discuss the physics of the parallel association in section \ref{subsec : discussion parallel}. The conservation laws for the device resulting from the parallel association are studied in Appendix~\ref{subsec : conservation laws parallel}. 

The parallel association of two TECs is represented on Fig. \ref{fig : parallel}.
Following the pin ensemble definitions given in Ref.~\cite{Raux2024} for the parallel association, the pins of devices 1, 2 and 3 are
\begin{align}
\mathscr{P}^{(1)}&=\lbrace E_1, C_1, E_2, C_2 \rbrace,\\
\mathscr{P}^{(2)}&=\lbrace E_3, C_3, E_4, C_4 \rbrace,\\
\mathscr{P}^{(3)}&=\lbrace \lbrace E_1, E_3\rbrace, \lbrace C_1, C_3\rbrace, \lbrace E_2, E_4 \rbrace, \lbrace C_2, C_4\rbrace \rbrace,
\end{align}
where $E_n$ (resp. $C_n$) is the index of the pin number $n$ associated to energy transport (resp. charge transport). We recall that the pins in $\mathscr{P}^{(3)}$ are called lumped pins. Each pin $P$ in $\mathscr{P}^{(3)}$ gathers the pins in $\mathscr{P}^{(1)}$ and $\mathscr{P}^{(2)}$ that are at the same local potential. This means that the local potentials at all pins of TEC 1 and 2 are functions of the potential on the lumped pins of TEC 3 as 
\begin{equation}
\bm a^{(m)}_\mathrm{i}=\bm \pi_\parallel^{(m,3)}\bm a^{(3)}_\mathrm{i},
\end{equation}
for $m=1,2$. The local potentials $\bm a^{(m)}_{\mathrm{i}}$ are defined in Eq.~\eqref{eq : physical forces conjugated to the internal currents} when adding superscript $(m)$ to $T_{\chi}$ and $V_{\chi}$ for $\chi = l,r$. The matrices $\bm \pi_\parallel^{(m,3)}=\mathbb{1}_{4}$ 
%
encode the topology of the parallel association of TEC 1 and 2. Given the symmetry of the association considered here they happen to be identical and equal to identity: TEC 1, 2 and 3 see the same potentials on their left and on their right.
\subsubsection{Equivalent conductance matrix}
\label{subsec : equivalent conductance matrix parallel}
We now turn to the determination of the conductance matrix for TEC 3. We have shown 
that the conductance matrix at physical level for the internal currents of a single TEC (here for $m=1,2$) reads
\begin{equation}
\bm g^{(m)}_i=
\begin{bmatrix}
\bm G^{(m)}_i & -\bm G^{(m)}_i\\
-\bm G^{(m)}_i & \bm G^{(m)}_i
\end{bmatrix}
\end{equation}
where $\bm G^{(m)}_i$ is the conductance matrix recalled in Eq.~\eqref{eq : global flux force relation conjugated}. Now, using the law of conductance matrix addition derived in Ref.~\cite{Raux2024}, the conductance matrix for device $3$ reads
\begin{equation}
\bm g^{(3)}=\sum_{m=1}^2\bm \pi_\parallel^{(m,3)T}\bm g^{(m)}\bm \pi_\parallel^{(m,3)}
\end{equation}
that simplifies into
\begin{equation}
\bm g^{(3)}=\begin{bmatrix}
\bm G^{(1)}+\bm G^{(2)} & - \bm G^{(1)}-\bm G^{(2)} \\
-\bm G^{(1)}-\bm G^{(2)} & \bm G^{(1)}+\bm G^{(2)}
\end{bmatrix}.
\end{equation}
%
The conductance matrix $\bm G^{(3)}$ for a set of linearly independent currents of devices 3, reads
\begin{equation}
\bm G^{(3)}=\bm S^{(3)+}\bm g^{(3)}\bm S^{(3)+T}
\end{equation}
where $\bm S^{(3)}$ is the selection matrix that selects a set of independent currents among the redundant currents flowing through the pins in $\mathscr{P}^{(3)}$, see Section~\ref{fundacurrentforce}. 
\subsubsection{Discussion}
\label{subsec : discussion parallel}
Insofar as the parallel configuration preserves the Dirichlet conditions for each of the two devices, the electric linear behaviors are also preserved by simple application of the superposition theorem. It is therefore natural that TEC 3 should also behave in a perfectly linear fashion. No impedance matching problem arise here as there is no potential continuity issues. But this observation does not exhaust the subject. Indeed, the parallel configuration requires that the two electric parts of the two devices be connected in parallel. In this case, a loop current flows between the two electric parts of the TECs, unless they each have the same voltage across their terminals. It's easy to see that the respective currents are given by: 
 \begin{eqnarray}
 i_{C_1}&=&-\frac{(\alpha^{(2)}-\alpha^{(1)}) \Delta T^{(3)}}{R^{(1)}+R^{(2)}}+\frac{R^{(2)}i_{\lbrace C_1,C_3\rbrace}}{R^{(1)}+R^{(2)}}, \\
 i_{C_3}&=&\frac{R^{(1)}i_{\lbrace C_1,C_3\rbrace}}{ R^{(1)}+R^{(2)} }+\frac{(\alpha^{(2)}-\alpha^{(1)}) \Delta T^{(3)}}{R^{(1)}+R^{(2)}}.
\end{eqnarray}
We can see that even when the current $i_{\lbrace C_1,C_3 \rbrace}$ is zero, an electric current flows through both devices, generating residual internal dissipation. From these two equations it is clear that one of the devices will act as an electrical generator, feeding the other acting as a receptor. 
Once again, it is the presence of Seebeck coefficient difference between TECs 1 and 2 that leads to an unconventional situation. However, it's important to note that this time it's not a signature of the Peltier effect, but simply a pure Joule dissipation, unlike in the previous serial case.

\section{Conclusion}

Starting from Ioffe's integrated current-force characteristics of TEC, we summarized the energy/matter current-force characteristics by introducing a conductance matrix relating the currents to their conjugated forces in the entropy production rate. This matrix is not unique since it corresponds to a choice of model for the coupling between matter and energy. Moreover, we gave an algebraic procedure to change the current basis using the conservation laws of the TEC. We have shown that two equivalent sets of currents are available for describing a TEC: the internal currents (energy and matter), and the external currents that can be easily determined by measuring the currents (heat and matter) entering the boundaries of the TEC. We determined the non-equilibrium conductance matrix for each current set. We found that nonequilibrium conductance matrices are in general force-dependent and apply far from equilibrium.

Then, we applied our theory of thermodynamic circuits to the serial and parallel associations of TECs. The law of resistance/conductance matrix addition takes, in this case, a simple form because of the symmetry of the associations and the small number of pins. It thus paves the way for a systematic study of thermoelectric networks since our toolbox is not limited to binary associations of TECs but can also be applied to more complex networks. 

Our study unveiled, in particular, that the serial association of two TECs described by the CPM follows an effective model akin to the CPM but with electric current-dependent thermoelectric coefficients, see Appendix \ref{effective model}. This emerges from non-ideal boundary conditions, i.e., from the electric current dependency of the local potentials at their interface. The law of resistance addition (serial association) gives the same (nonlinear) current-force characteristics for TEC 3 as the one of the effective model. However, the coupling between the currents is only predicted by the law of resistance addition. It is not in other approaches, preventing us from validating this part of our circuit theory on such thermodynamic models of TECs. In this regard, further investigations with stochastic modeling are required. We achieve a step in this direction in Ref.~\cite{Raux2024c} dealing with chemical reaction networks. 
Finally, we studied the role of the in-homogeneity of thermoelectric coefficients by comparing the spatial profiles of temperature, electric potentials and the two composite potentials from which derive electric and energy currents. The mismatch between thermoelectric coefficients significantly modify these profiles rising question of impedance adaptation beyond the scalar case. 

In any case, it would be interesting to investigate the link between the non-equilibrium conductance matrix of a TEC and the fluctuations of its currents in light of the results on Markov jump processes derived in Ref.~\cite{Vroylandt2018}. In this work, the authors show that the non-equilibrium conductance matrix gives the best lower bound on the current covariance. Our circuit theory would then allow us to determine the best lower bound on the current covariances of a composite TEC, given the bounds on the current covariances of the two subdevices.

\section*{Acknowledgments}
P.R. acknowledges fruitful discussions with Armand Despons.

\appendix

\section{Structure of the selection matrix}
\label{sec : structure of the selection matrix}
In this appendix, we justify the form of the selection matrix given in Eq.~\eqref{eq : structure of S}. The matrix of conservation law $\bm \ell $ is of dimension $|\mathscr{L}| \times |\mathscr{P}|$ and $\mathrm{rank}(\bm \ell)=|\mathscr{L}| $. It describes the $|\mathscr{L}|$ linear relations between the physical currents that we write $\bm \ell \bm i = 0$. Given the rank of $\bm \ell$, we can chose the order of $\bm i$'s components such that the $|\mathscr{L}|$ last columns of matrix $\bm \ell $, denoted $\bm \ell_{d}$, are linearly independent. The remaining $|\mathscr{I}|$ first columns of $\bm \ell$ are denoted $\bm \ell_{I}$. Then, we take the $|\mathscr{I}|$ first components of $\bm i$ as our choice of fundamental currents such that
\begin{equation}
\bm i =
\begin{pmatrix}
\bm I\\
\bm i_\mathrm{d}
\end{pmatrix} \quad \text{ and } \quad 
\bm \ell = 
\begin{pmatrix}
\bm \ell_I & \bm \ell_{d}
\end{pmatrix}.
\label{eq : splitting}
\end{equation}
The last $|\mathscr{L}|$ components of $\bm i$, denoted $\bm i_{d}$, are the dependent currents that can be obtained as linear combination of the components of $\bm I$. Then, the square matrix $\bm \ell_{d}$ has rank equal to its dimension $|\mathscr{L}|$ and hence is invertible. By construction, the selection matrix relating physical and fundamental currents by $\bm i= \bm S \bm I$ writes
\begin{equation}
\bm S = \begin{bmatrix}
\mathbbm{1} \\
\bm T
\end{bmatrix}, \label{struc selection matrix}
\end{equation}
where $\mathbbm{1}$ is the identity matrix of dimension $|\mathscr{I}|$ and the matrix $\bm T$ remains to be determined using  
\begin{equation}
\bm 0 = \bm \ell \bm S = \bm\ell_I + \bm\ell_{d} \bm T = 0 \quad \Rightarrow \quad \bm T = - \bm \ell_{d}^{-1} \bm \ell_{I}.
\end{equation}
This justifies the form of selection matrix given in Eq.~\eqref{eq : structure of S}.
\section{Non unicity of the conductance matrices}
\label{sec : non unicity of the conductance matrices}
Exhibiting a conductance matrix from the current-force relation cannot be done uniquely without relying on a microscopic model. We illustrate this fact for our model of TEC, starting from the non equilibrium conductance matrix of Eq.~\eqref{eq : global flux force relation conjugated}. 
Let's assume that there exists a symmetric matrix $\bm G_\mathrm{i}^{'}\neq \bm G_\mathrm{i}$ such that
\begin{equation}
\bm G_\mathrm{i}^{'} = \frac{T_lT_r}{R\bar{T}}\begin{bmatrix}
a & b \\
b & c
\end{bmatrix},
\end{equation}
with the same current-force characteristics
\begin{equation}
\bm I_\mathrm{i}=\bm G_\mathrm{i} \bm A_\mathrm{i} = \bm G_\mathrm{i}^{'}\bm A_\mathrm{i}.
\label{eq : flux equality}
\end{equation}
This relation writes
\begin{equation}
\begin{bmatrix}
KR\bar{T} + F^2  & F\\
F & 1
\end{bmatrix}
\begin{pmatrix}
A_E\\
A_{C}
\end{pmatrix}=
\begin{bmatrix}
a & b\\
b & c
\end{bmatrix}
\begin{pmatrix}
A_E\\
A_{C}
\end{pmatrix}.
\end{equation}
Each line of this equation can be solved for $b$ as function of $a$ and $c$. Enforcing the equality of the resulting $b$ constrains $a$ and $c$ by
\begin{equation}
\frac{A_E}{A_{C}}\left(KR\bar{T} + F^2 - a \right) = \frac{A_{C}}{A_E}\left( 1-c \right).
\end{equation}
As a consequence, $\bm G_\mathrm{i}^{'}$ writes
\begin{equation}
\bm G_\mathrm{i}^{'}=\bm G_\mathrm{i}
+ \frac{T_lT_r}{R\bar{T}}
\begin{bmatrix}
-\left(\frac{A_{C}}{A_E}\right)^2 & \frac{A_{C}}{A_E}\\
\frac{A_{C}}{A_E} & -1
\end{bmatrix}
(1-c), \label{eq: equiv class}
\end{equation}
showing that the nonequilibrium conductance matrix $\bm G_\mathrm{i}$ is not unique when providing the current-force relation only, with no information on the currents fluctuations. Any value of $c$ in the above equation produces a conductance matrix compatible with this current-force relation. We emphasize that $c$ is in principle a function of $A_E$ and $A_{C}$. Close-to-equilibrium, the non-equilibrium conductance coincides by definition with Onsager's response matrix: the knowledge of the currents covariance close to an equilibrium states allows to define the conductance matrix uniquely.
The fact that a non-equilibrium conductance matrix includes more information on the model than the current-force relation is argued in Ref.~\cite{Vroylandt2018}. A non-equilibrium conductance matrix can arise (with a unique definition) from a microscopic modeling. Another approach is to consider it as an alternative way of defining a model, with the idea that the additional information present in the matrix constrains the quadratic fluctuations of currents whatever the considered non-equilibrium stationary state.

\section{Serial association}
In this Appendix, we apply the general method of Ref.~\cite{Raux2024} to obtain the conservation laws for TEC 3 in section \ref{subsec : conservation laws} and to justify the equivalent resistance matrix of Eq.\eqref{eq : R1+R2} from the law of resistance matrix addition in section \ref{conductance matrix dimension matching}. We obtain the same characteristic equations by a direct approach in section \ref{effective model}.

\subsection{Conservation laws}
\label{subsec : conservation laws}
Let's consider the internal physical currents as given in Eq.~\eqref{internalcurrents}, but specified for TEC $m=1,2,3$ :
\begin{equation}
\bm i^{(m)}=
\begin{pmatrix}
	\bm i_{l}^{(m)} \\ \bm i_{r}^{(m)}
\end{pmatrix} \quad \text{with } \bm i_{\chi}^{(m)} = \begin{pmatrix}
i_{E\chi}^{(m)}\\
i_{C\chi}^{(m)}
\end{pmatrix} \text{ for } \chi= l,r. \label{eq : im}
\end{equation}
Conservation laws in matrix form write for $m=1,2$
\begin{equation}
\bm \ell^{(m)}\bm i^{(m)}=0,
\label{eq : conservation laws 1 and 2 (1)}
\end{equation}
with
\begin{equation}
\bm{\ell}^{(m)}=
\begin{bmatrix}
\mathbb{1}_2 & \mathbb{1}_2
\end{bmatrix},
\label{eq : conservation laws 1 and 2 (2)}
\end{equation}
where $\mathbb{1}_2$ is the $2\times 2$ identity matrix. We expect identical conservation laws for TEC 3 given its similarity with TEC 1 and 2. As an illustration, we recover below this trivial result by using the method of Ref.~\cite{Raux2024}. 

Conservation of energy and charge at the interface leads to
\begin{equation}
\bm L_e \bm i^{(3)}=\bm L_i \bm i_r^{(1)}
\label{eq : lois de conservation enesemble 2}
\end{equation}
where we define
\begin{equation}
\bm L_e=\begin{bmatrix}
\mathbb{1}_2 & \mathbb{0}_2\\
\mathbb{0}_2 &  \mathbb{1}_2
\end{bmatrix},
\quad 
\bm L_i=
\begin{bmatrix}
- \mathbb{1}_2\\
 \mathbb{1}_2 
\end{bmatrix},
\label{eq : lois de conservation concéténées}
\end{equation}
with $\mathbb{0}_{2}$ the $2\times 2$ null matrix.
The conservation laws for TEC $3$ are obtained by multiplying Eq.~\eqref{eq : lois de conservation enesemble 2} on the left by $\bm v$, the matrix whose lines are the basis vector of $\mathrm{coker}(\bm L_i)$ such that $\bm v \bm L_i=\bm 0$. This matrix $\bm v$ reads here
\begin{equation}
\bm v= \begin{bmatrix}
\mathbb{1}_2 & \mathbb{1}_2
\end{bmatrix}.
\end{equation}
Matrix of conservation laws for TEC 3 is as anticipated
\begin{equation}
\bm \ell^{(3)}=\bm v \bm L_e=\begin{bmatrix}
\mathbb{1}_2 & \mathbb{1}_2
\end{bmatrix}.
\end{equation}

\subsection{Conductance matrix dimension matching}
\label{conductance matrix dimension matching}
For each devices $m=1,2,3$, the vector of linearly dependent currents $\bm i^{(m)}$ follows from the vector of independent currents $\bm I^{(m)}$ as 
\begin{equation}
\bm i^{(m)}=\bm S^{(m)}\bm I^{(m)}
\end{equation}
where $\bm S^{(m)}$ is a selection matrix whose column vectors are chosen as basis vectors of $\mathrm{ker}(\bm \ell^{(m)})$ such that $\bm \ell^{(m)}\bm S^{(m)}=\bm 0$. For all $m=1,2,3$, we choose the following vector of linearly independent currents
\begin{equation}
\bm I^{(m)}=
\begin{pmatrix}
i_{El}^{(m)}\\
i_{Cl}^{(m)}
\end{pmatrix}=\begin{pmatrix}
I_E\\
I_C
\end{pmatrix}
\end{equation}
which is associated to the following choice of selection matrix and its associated pseudoinverse
\begin{equation}
\bm S^{(m)}=
\begin{bmatrix}
\mathbb{1}_2\\
-\mathbb{1}_2
\end{bmatrix}\quad \mathrm{and}\quad {\bm S^{(m)}}^+=\frac{1}{2}\begin{bmatrix}
\mathbb{1}_2 & -\mathbb{1}_2
\end{bmatrix}.
\end{equation}
Eqs.~(\ref{eq : global flux force relation conjugated}--\ref{eq : global flux force relation conjugated}) written for device $m=1,2$ can be inverted
\begin{equation}
\bm A^{(m)}=\bm R^{(m)}\bm I^{(m)}
\label{eq : current force fundamental}
\end{equation}
with 
\begin{eqnarray}
\!\!\!\bm R^{(m)}=&& \nonumber
\frac{1}{K^{(m)}T_l^{(m)}T_r^{(m)}} \\ &&
\begin{bmatrix}
1 & -F^{(m)}\\
-F^{(m)} & K^{(m)}R^{(m)}\bar{T}^{(m)}+{F^{(m)}}^2 
\end{bmatrix}.
\label{eq : conductance matrix m 1 et 2}
\end{eqnarray}
and
\begin{equation}
\bm A^{(m)}=\begin{pmatrix}
A_E^{(m)}\\
A_C^{(m)}
\end{pmatrix}=
\begin{pmatrix}
\frac{1}{T_r^{(m)}}-\frac{1}{T_l^{(m)}}\\
\frac{V_l^{(m)}}{T_l^{(m)}}-\frac{V_r^{(m)}}{T_r^{(m)}}
\end{pmatrix}
\end{equation}
where $A_E^{(m)}$ (resp. $A_C^{(m)}$) is the affinity conjugated to the energy (resp. charge) current in the EPR. Following Ref.~\cite{Raux2024} to determine $\bm G^{(3)}$, we start solving Eq.~\eqref{eq : lois de conservation enesemble 2} for $\bm i_{r}^{(1)}$ which yields 
\begin{equation}
\bm i_{r}^{(1)}=\frac{1}{2}\begin{bmatrix}
-\mathbb{1}_2 & \mathbb{1}_2
\end{bmatrix} \bm i^{(3)}
\end{equation}
where we used the left pseudo-inverse of $\bm L_i$ since its columns  are linearly independent. This last relation can thus be used to express $\bm i^{(m)}$ for $m=1,2$ in term of $\bm i^{(3)}$ as 
\begin{equation}
\bm i^{(m)}=\bm \pi^{(m,3)}\bm i^{(3)},
\end{equation}
where 
\begin{equation}
\bm \pi^{(1,3)}=
\begin{bmatrix}
\mathbb{1}_2 & \mathbb{0}_2\\
-\frac{1}{2}\mathbb{1}_2 & \frac{1}{2}\mathbb{1}_2
\end{bmatrix},
\quad 
\bm \pi^{(2,3)}=
\begin{bmatrix}
\frac{1}{2}\mathbb{1}_2 & -\frac{1}{2}\mathbb{1}_2\\
\mathbb{0}_2 & \mathbb{1}_2
\end{bmatrix}.
\end{equation}
Now the law of resistance addition reads 
\begin{equation}
\bm R^{(3)}=\sum_{m=1}^2 \bm\Pi^{(m,3)T}\bm R^{(m)}\bm \Pi^{(m,3)}
\label{eq : law of resistance matrix addition}
\end{equation}
where 
\begin{equation}
\bm \Pi^{(m,3)}\equiv\bm S^{(m)+}\bm \pi^{(m,3)}\bm S^{(3)}=\mathbb{1}_2 .
\end{equation}
We thus obtain the result stated in the main text
\begin{equation}
\bm R^{(3)}=\bm R^{(1)}+\bm R^{(2)}.
\label{eq : law of resistance matrix addition 2}
\end{equation}

\subsection{Effective model}
\label{effective model}

We show here that TEC 3 can be modelled by an effective model akin to the CPM, but with $I_C$ dependant thermoelectric coefficients. Let's consider the following averages and differences of thermoelectric coefficients for TEC 1 and 2
\begin{align}
\bar{\alpha}&=\frac{\alpha^{(1)}+\alpha^{(2)}}{2}, & \delta_\alpha&=\alpha^{(2)}-\alpha^{(1)},\\
\bar{K}&=\frac{K^{(1)}+K^{(2)}}{2}, & \delta_K&=K^{(2)}-K^{(1)},\\
\bar{R}&=\frac{R^{(1)}+R^{(2)}}{2}, & \delta_R&=R^{(2)}-R^{(1)},
\end{align}
which, combined with Eqs.~(\ref{eq : global matter current},\ref{eq : global energy current}), yields 
\begin{align}
    I_E&=\frac{I_E^{(1)}+I_{E}^{(2)}}{2}=-K^{(3)}\Delta T^{(3)}+F^{(3)}I_C,\label{eq : ie3}\\
    I_C&=\frac{I_C^{(1)}+I_C^{(2)}}{2}=-\frac{\alpha^{(3)}\Delta T^{(3)}+\Delta V^{(3)}}{R^{(3)}},\label{eq : ic3}
\end{align}
with the following thermoelectric coefficients for TEC 3
\begin{align}
    \alpha^{(3)}&=\bar{\alpha}-\frac{\delta_\alpha}{2(2\bar{K}+\delta_\alpha I_C)} \label{eq : alpha3},\\
    K^{(3)}&=\frac{\bar{K}}{2}-\frac{\delta_K \delta_\alpha}{4(2\bar{K}+\delta_\alpha I_C)} \label{eq : K3},\\
    R^{(3)}&=2\bar{R}+\delta_\alpha \frac{\delta_\alpha \bar{T}^{(3)}-\bar{R}I_C}{2\bar{K}+\delta_{\alpha} I_C}\label{eq : R3},\\
    F^{(3)}&=\frac{F^{(1)}+F^{(2)}}{2}-\frac{\delta_K(\delta_\alpha \bar{T}^{(3)}-\bar{R}I_C)}{2(2\bar{K}+\delta_{\alpha} I_C)}. \label{eq : F3}
\end{align}
Since we consider an imposed $\Delta T ^{(3)}$ value, we see that all these four parameters now depend on the value of the electric current $I_C$. According to Eq.~\eqref{eq : global flux force relation conjugated}, the current--force characteristics of TEC 3 can thus be described by the resistance matrix $\bm R^{(3)'}$ which reads
\begin{equation}
    \bm R^{(3)'}=\frac{1}{K^{(3)}T_lT_r}\begin{bmatrix}
    1 & -F^{(3)}\\
    -F^{(3)} & K^{(3)}R^{(3)}\bar{T}^{(3)}+{F^{(3)}}^2
    \end{bmatrix}
\end{equation}
for the force and current basis of Eq.~\eqref{eq : force current device 3}. Clearly, this last matrix does not coincides with matrix $\bm R^{(3)}$ obtained by the law of resistance addition. However, since $\bm R^{(3)'}\bm I=\bm R^{(3)}\bm I$ by construction, these two matrices belong to the same class of resistance matrix. This can be stated with an equivalent of Eq.~\eqref{eq: equiv class}, but for resistance matrices:
\begin{equation}
    \bm R^{(3)'}=\bm R^{(3)}+ \left( R^{(3)'}_{22}- R^{(3)}_{22} \right)
    \begin{bmatrix}
    -\left(\frac{I_C}{I_E}\right)^2 & \frac{I_C}{I_E}\\
    \frac{I_C}{I_E} & 1
    \end{bmatrix},
\end{equation}
with
\begin{equation}
    \begin{split}
    R^{(3)'}_{22}- R^{(3)}_{22}
    &=\frac{1}{K^{(3)}}\left[\frac{{F^{(1)}}^2+{F^{(2)}}^2}{2T_lT_r} - \frac{1}{T} \left( \frac{{F^{(1)}}^2}{T_l}+\frac{{F^{(2)}}^2}{T_r}\right) \right. \\
    &+ \left.\frac{{F^{(1)}}{F^{(2)}}}{T_lT_r}\right] + \frac{R^{(3)}}{2}\left(A_E^{(2)} - A_E^{(1)}\right),
\end{split}
\end{equation}
and $T$ the temperature at the interface given in Eq.~\eqref{eq : T}.

\section{Parallel association: conservation laws}
\label{subsec : conservation laws parallel}
We determine $\bm \ell^{(3)}$ in the case of a parallel association following the method of Ref.~\cite{Raux2024}. Eq.~\eqref{eq : conservation laws 1 and 2 (2)} provides the conservation of internal currents for TEC 1 and 2, i.e., their conservation laws before parallel association. The external conservation laws, i.e. the conservation laws arising from the parallel association of TEC 1 and 2 to create TEC 3, are
\begin{equation}
\bm i^{(3)}=
\begin{bmatrix}
\bm \pi_\parallel^{(1,3)T} & \bm \pi_\parallel^{(2,3)T}  
\end{bmatrix}
\begin{pmatrix}
\bm i^{(1)}\\
\bm i^{(2)}
\end{pmatrix},
\end{equation}
where $\bm \pi^{(m,3)} = \mathbb{1}_{4}$ for $m=1,2$ and
\begin{equation}
\bm i^{(3)}=
\begin{pmatrix}
i_{\displaystyle \lbrace E_1, E_3\rbrace}\\
i_{\displaystyle\lbrace C_1, C_3\rbrace}\\
i_{\displaystyle\lbrace E_2, E_4 \rbrace}\\
i_{\displaystyle \lbrace C_2, C_4\rbrace }\\
\end{pmatrix}, \quad 
\begin{pmatrix}
\bm i^{(1)}\\
\bm i^{(2)}
\end{pmatrix}=\displaystyle
\begin{pmatrix}
i_{\displaystyle E_1}\\
i_{\displaystyle C_1}\\
i_{\displaystyle E_2}\\
i_{\displaystyle C_2}\\
i_{\displaystyle E_3}\\
i_{\displaystyle C_3}\\
i_{\displaystyle E_4}\\
i_{\displaystyle C_4}
\end{pmatrix}.
\end{equation}
We combine all the conservation laws into
\begin{equation}
\bm L \begin{pmatrix}
\bm i^{(1)}\\
\bm i^{(2)}
\end{pmatrix} = 
\begin{pmatrix}
0\\
0\\
0\\
0\\
\bm i^{(3)}
\end{pmatrix},
\label{eq : conservation law parallel (1)}
\end{equation}
with
\begin{equation}
\bm L= \begin{bmatrix}
\bm \ell^{(1)} & \bm 0\\
0 & \bm \ell^{(2)}\\
\bm \pi_\parallel^{(1,3)T} & \bm \pi_\parallel^{(2,3)T}
\end{bmatrix} 
 = 
\left[
\begin{array}{cccc|cccc}
1 & 0 & 1 & 0 & 0 & 0 & 0 & 0\\
0 & 1 & 0 & 1 & 0 & 0 & 0 & 0\\
\hline
0 & 0 & 0 & 0 & 1 & 0 & 1 & 0\\
0 & 0 & 0 & 0 & 0 & 1 & 0 & 1\\
\hline
1 & 0 & 0 & 0 & 1 & 0 & 0 & 0\\
0 & 1 & 0 & 0 & 0 & 1 & 0 & 0\\
0 & 0 & 1 & 0 & 0 & 0 & 1 & 0\\
0 & 0 & 0 & 1 & 0 & 0 & 0 & 1
\end{array}
\right].
\end{equation} 
A left null basis of $\bm L$ reads 
\begin{equation}
\bm u= 
\begin{bmatrix}
-1 & 0 & -1 & 0 & 1 & 0 & 1 & 0\\
0 & -1 & 0 & -1 & 0 & 1 & 0 & 1
\end{bmatrix}.
\end{equation}
We have shown in Ref.~\cite{Raux2024} that the last $|\mathscr{P}^{(3)}|=4$ columns of $\bm u$ gives the matrix of conservation laws for TEC 3, here 
\begin{equation}
\bm \ell^{(3)}=
\begin{bmatrix}
1 & 0 & 1 & 0\\
0 & 1 & 0 & 1
\end{bmatrix}.
\end{equation} 
Indeed, left multiplying  Eq.~\eqref{eq : conservation law parallel (1)} by $\bm u$ gives a null vector on the left hand side, and the right hand side exhibits the linear dependence between the components of $\bm i^{(3)}$, i.e. between the physical current of TEC 3. Here again, TEC 1, 2 and 3 all have the same conservation laws, which was expected by symmetry.

\bibliographystyle{unsrt}

\bibliography{bibliocircuitation}

\end{document}